\documentclass[%
 reprint,
 amsmath,amssymb,
 aps,
]{revtex4-1}

\usepackage{graphicx}
\usepackage{dcolumn}
\usepackage{bm}
\usepackage[colorlinks, linkcolor=blue, anchorcolor=blue, citecolor=blue]{hyperref}
\usepackage{url}
\usepackage{amsmath,amssymb}
\usepackage{newtxmath}
\usepackage[mathlines]{lineno}
\usepackage{subfigure}
\usepackage{eurosym}

\begin{document}

\preprint{APS/123-QED}

\title{Manipulating single-photon transport in a waveguide-QED structure containing two giant atoms}
\author{S. L. Feng}
\affiliation{School of Physical Science and Technology, Southwest Jiaotong University, Chengdu 610031, China}
\author{W. Z. Jia}
\email{wenzjia@swjtu.edu.cn}
\affiliation{School of Physical Science and Technology, Southwest Jiaotong University, Chengdu 610031, China}
\date{\today}

\begin{abstract}
We investigate coherent single-photon transport in a waveguide-QED structure containing two giant 
atoms. The unified analytical expressions of the single-photon scattering amplitudes applicable for 
different topological configurations are derived. The spectroscopic characteristics in different 
parameter regimes, especially the asymmetric  Fano lineshapes and the EIT-like spectra, are 
analyzed in detail. Specifically, we find that the appearance of Fano lineshapes is influenced by not only the phase 
delays between coupling points but also the topologies of system. We also summarize the general 
conditions for appearance of EIT-like spectra by analyzing the master equation, 
and verify these conditions by checking the corresponding analytical expressions of the scattering spectra. These phenomena may provide powerful tools for controlling and manipulating 
photon transport in the future quantum networks. 
\end{abstract}

\pacs{Valid PACS appear here}
\maketitle


\section{\label{Introduction}Introduction}
The realization of light-matter interactions at the single-photon level plays a central role in the fields
of modern quantum optics and quantum information processing. This goal can be realized by 
strongly coupling a single atom or multiple atoms to a one-dimensional (1D) waveguide, called 
waveguide quantum electrodynamics (wQED) system \cite{Roy-PRM2017, Gu-PhysReports2017}. These 
kinds of structures exhibit high atom-waveguide coupling efficiency, resulting in low leakage of 
photons into unguided degrees of freedom \cite{Astafiev-Sci2010, Hoi-PRL2011}. 
This feature makes the wQED systems become excellent platforms to manipulate transport of single 
or few photons \cite{Astafiev-Sci2010,Hoi-PRL2011,Shen-Opt.Lett.2005,Shen-PRL2005,Chang-PRL2006,Shen-PRL2007,Chang-Nature2007,Zhou-PRL2008,Shi-PRB2009,Shen-PRA2009,Longo-PRL2010,Zheng-PRA2010,Fan-PRA2010,Witthaut-NJP2010,Roy-PRL2011,Zheng-PRL2011,Hoi-PRL2013,Jia-PRA2013,Laakso-PRL2014,Yang-Ann.Phys.(Berlin)2020}. Thus quantum devices with high efficiency, including quantum routers \cite{Abdumalikov-PRL2010,Hoi-PRL2011,Bermel-PRA2006,Zhou-PRL2008,Aoki-PRL2009,Zhu-PRA2019}, single-photon transistors \cite{Chang-Nature2007,Neumeier-PRL2013} and quantum frequency converters \cite{Bradford-PRL2012,Bradford-PRA2012,Wang-PRA2014,Zhao-PRA2017,Jia-PRA2017,Cao-JPB2021}, can be 
realized in the wQED structures.  In particular, when two or more atoms coupled to a 1D continuum, the 
interactions mediated by the guided modes as well as the interferences between photons re-emitted 
by different atoms, can enable a number of interesting effects, such as asymmetric Fano lineshapes \cite{Tsoi-PRA2008,Cheng-OSA2012,Liao-PRA2015,Cheng-PRA2017,Mukhopadhyay-PRA2019}, 
electromagnetically induced transparency (EIT) without control field \cite{Shen-PRB2007,Fang-PRA2017,Mukhopadhyay-PRA2020}, waveguide-mediated entanglement between distant atoms \cite{Zheng-PRL2013,Ballestero-PRA2014,Facchi-PRA2016,Mirza-PRA2016},  generation of photonic band gap \cite{Fang-PRA2015,Greenberg-PRA2021}, cavity QED with atomic mirrors \cite{Chang-IOP2012,Mirhosseini-Nature2019}, creating and engineering superradiant and subradiant states \cite{Loo-Sci2013,Zhang-PRL2019,Ke-PRL2019,Wang-PRL2020}, and so on. 

Recently, with the development of modern nanotechnology, a new type of wQED structure containing 
the so called giant atoms has brought about widespread attention 
\cite{Kockum-MI2020,Kockum-PRA2014}.
In these setups, artificial atoms (e.g., transmon qubits \cite{Koch-PRA2007}) can couple to the 
bosonic modes (phonons or microwave photons) in a 1D waveguide at multiple points being spaced 
wavelength distances apart. Different from the usual wQED structures with point-like small atoms, 
the multiple coupling points of a single giant atom can provide additional interference effects, 
resulting in some novel phenomena, such as frequency-dependent decay rate and Lamb shift of a 
single atom \cite{Kockum-PRA2014,Kannan-Nature2020,Vadiraj-PRA2021,Yu-PRA2021}, and 
decoherence-free interaction between two braided giant atoms \cite{Kockum-PRL2018,Kannan-Nature2020}. By utilizing non-Markovianity originated from the time-delay between coupling points, a 
giant atom can realize polynomial spontaneous decay \cite{Guo-PRA2017,Andersson-Nature2019}, 
and create bound states of bosons \cite{Guo-Phys.Rev.Research2020,Zhao-PRA2020,Guo-PRA2020}.  

In this paper, we focus on the single-photon transport properties in a 1D waveguide coupled by two 
giant atoms. Utilizing a real-space scattering method \cite{Shen-PRL2005}, we obtain the general 
analytical expressions of the single-photon scattering amplitudes, which are available for three 
basic topologies \cite{Kockum-PRL2018} of the double-giant-atom systems. It is shown that the 
scattering spectra are determined by
the following characteristic quantities of the system:  the Lamb shift, the individual decay, the exchange 
interaction, and the collective decay. Then, based on these general expressions, we further analyze two 
important phenomena, asymmetric Fano lineshape and EIT without control field, which can be 
observed from the scattering spectra. These phenomena also exist in the wQED system containing multiple
small atoms \cite{Tsoi-PRA2008,Cheng-OSA2012,Liao-PRA2015,Cheng-PRA2017,Mukhopadhyay-PRA2019,Shen-PRB2007,Fang-PRA2017,Mukhopadhyay-PRA2020}, but their counter part in giant-atom systems will give rise to some different features due
to additional interference effects and diverse configurations. Specifically, we find that the appearance of Fano interferences 
is strongly influenced by both the phase delay between coupling points and the 
topologies of system. In addition, we show that the scattering spectrum is also a powerful tool for characterizing the 
light-matter interactions, e.g., the decoherence-free interactions, in giant-atom structures. 
On the other hand, the phenomenon of EIT without control field was also firstly investigated in  
the wQED system containing two small atoms \cite{Shen-PRB2007,Fang-PRA2017,Mukhopadhyay-PRA2020}, 
and very recently similar phenomenon in the double-giant-atom system for some special 
cases was discussed in Ref.~\cite{Ask-arXiv2020}. Compared with the well known EIT with natural 
atoms \cite{Harris-PRL1990,Harris-PhysicsToday-1997,Fleischhauer-Rev.Mod.Phys.2005,Boller-PRL1991}, the control-field-free scheme may 
provide alternative ways to produce EIT-type phenomenon in solid-state systems like 
superconducting circuits, which are
hard to use as standard $\Lambda$ systems. Here, by analyzing the master equation, we obtain the \textit{general} 
conditions for EIT without control field, which are appropriate for all configurations of 
double giant atoms. As a verification, we further derive the expressions of the EIT-like spectra under 
these conditions from the scattering method. Our analysis show that in a wQED system 
with two giant atoms, the bright and the dark states required by
EIT-like phenomenon can be either collective or single-atom states if the parameters are properly
chosen. The conditions given in our paper may provide good guidance for the future experiments on 
EIT without control field and on-chip photon manipulation.

The paper is organized as follows. In Sec.~\ref{Model}, we give a theoretical model, including the 
system Hamiltonian and corresponding equations of motion, and further obtain the general 
expressions of single-photon scattering amplitudes. In Sec.~\ref{Fano Interferences}, we discuss the 
phase-dependent Fano lineshapes in giant-atom systems with different topological configurations. In 
Sec.~\ref{EITwithoutCF}, we provide the general conditions to realize EIT without control field in  
wQED systems containing two giant atoms. Finally, further discussions and conclusions are given in 
Sec.~\ref{conclusion}
\section{\label{Model}Model and solutions}
\subsection{\label{HamiltonianEOM}Hamiltonian and equations of motion}

\begin{figure}[t]
\includegraphics[width=0.5\textwidth]{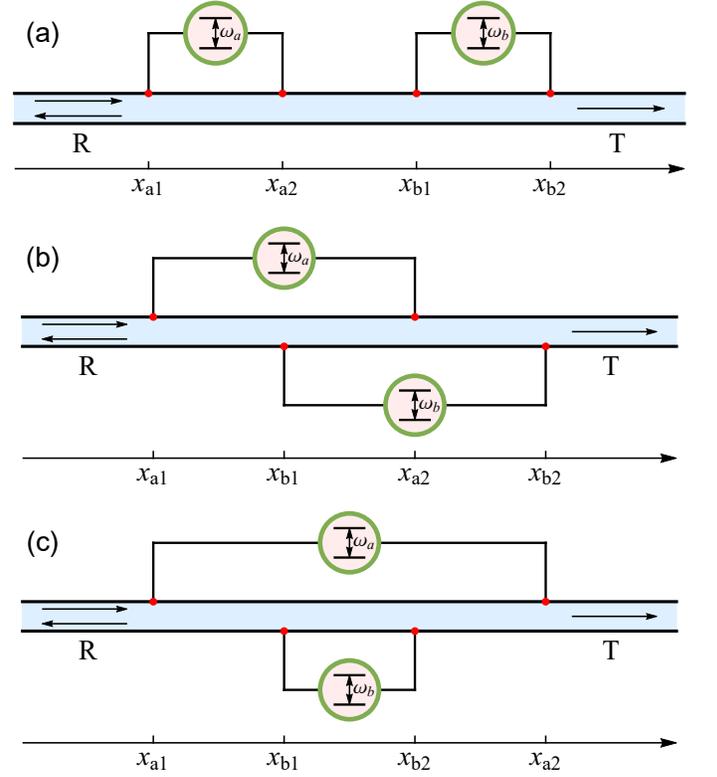}
\caption{Sketches of two giant atoms coupled to an open waveguide for three distinct topologies: (a) two separate giant atoms, (b) two braided giant atoms, (c) two nested giant atoms.}
\label{system}
\end{figure}

Here we focus on the wQED structures with two two-level giant atoms, and each atom 
couples to a 1D waveguide through two connection points. As summarized in Ref.~\cite{Kockum-PRL2018}, 
there are three different topologies for double-giant-atom wQED structures, called separate 
giant atoms, braided giant atoms, and nested giant atoms, respectively. The corresponding 
configurations are shown schematically in Figs.~\ref{system}(a)-\ref{system}(c). The atom with the leftmost 
coupling point is labelled by $a$ and the other by $b$.
The coordinates of the connecting points are $x_{jn}$, with $j=a,b$ labelling the 
atom and $n=1,2$ denoting the left and the right coupling points of each atom.
Under rotating-wave approximation (RWA), the Hamiltonian of the system can be written as $(\hbar=1)$
\begin{eqnarray}
\hat{H}&=&
\int\mathrm{d}x\hat{c}_{\mathrm{R}}^{\dagger}\left(x\right)\left(-iv_{\mathrm{g}}\frac{\partial}{\partial x}\right){\hat{c}_{\mathrm{R}}\left(x\right)}
\nonumber \\
&&+\int\mathrm{d}x\hat{c}_{\mathrm{L}}^{\dagger}\left(x\right)\left(iv_{\mathrm{g}}\frac{\partial}{\partial x}\right){\hat{c}_{\mathrm{L}}\left(x\right)}
+\sum_{j}\omega_{j}\hat{\sigma}_{j}^{+}\hat{\sigma}_{j}^{-}
\nonumber \\
&&+\sum_{s,j,n}\int\mathrm{d}x\ V_{jn}\delta\left(x-x_{jn}\right)\left[\hat{c}_{s}^{\dagger}(x)\hat{\sigma}_{j}^{-}+\mathrm{H.c.}\right],
\label{Hamiltonian}
 \end{eqnarray}
where $s=\mathrm{R},\mathrm{L}$, $j=a,b$, and $n=1,2$. $\hat{c}_{\mathrm{R}}^{\dagger}(x)$ $[\hat{c}_{\mathrm{R}}(x)]$ and $\hat{c}_{\mathrm{L}}^{\dagger}\left(x\right)$ $[\hat{c}_{\mathrm{L}}(x)]$ are the
field operators of creating (annihilating) the right- and
left-propagating photons at position $x$ in the waveguide. $\hat{\sigma}^{+}_{j}$ ($\hat{\sigma}^{-}_{j}$) is the raising (lowering) operator of the atom $j$. $v_{\mathrm{g}}$ is the group velocity of the photons in the waveguide. $\omega_{j}$ is the atomic transition frequency. $V_{jn}$ is the coupling strength at position $x_{jn}$.

We assume that initially a single photon with energy $\omega=v_{\mathrm{g}}k$ is incident from the left, where $k$ is the wave vector of the
photon. In the single excitation subspace, the interacting eigenstate of system can be written as
\begin{eqnarray}
\left|\Psi\right>&=&
\sum_{s}\int \mathrm{d}x\Phi_{s}\left(x\right)\hat{c}_{s}^{\dagger}\left(x\right)\left|\emptyset\right>
+\sum_{j}f_{j}\hat{\sigma}_{j}^{+}\left|\emptyset\right>,
\label{EigenState}
\end{eqnarray} 
where $\left|\emptyset\right>$ is the vacuum state, which means that there are no photons in the waveguide, and meanwhile the atoms are in their ground states. $\Phi_{s}(x)$ ($s=\mathrm{R}, \mathrm{L}$) is the single-photon wave function in the $s$ mode. $f_{j}$ ($j=a,b$) is the excitation amplitude of the atom $j$.
Substituting Eq.~\eqref{EigenState} into the eigen equation
\begin{equation}
\hat{H}\left|\Psi\right>=\omega\left|\Psi\right>
\label{eigenequation}
\end{equation}
yields the following equations of motion:
\begin{subequations}
\begin{align}
&\left(-iv_{\mathrm{g}}\frac{\partial}{\partial x}-\omega\right)\Phi_{\mathrm{R}}\left(x\right)
+\sum_{j,n}f_{j}V_{jn}\delta\left(x-x_{jn}\right)=0,
\label{motion1}
\end{align}
\begin{align}
&\left(iv_{\mathrm{g}}\frac{\partial}{\partial x}-\omega\right)\Phi_{\mathrm{L}}\left(x\right)
+\sum_{j,n}f_{j}V_{jn}\delta\left(x-x_{jn}\right)=0,
\label{motion2}
\end{align}
\begin{align}
&\left(\omega_{j}-\omega\right)f_{j}+\sum_{s,n}V_{jn}\Phi_{s}\left(x_{jn}\right)=0.
\label{motion3}
\end{align}
\end{subequations}
\subsection{\label{GeneralSolutions}General expressions of the scattering amplitudes}
For a photon incident from the left, $\Phi_{\mathrm{R}}\left(x\right)$ and $\Phi_{\mathrm{L}}\left(x\right)$ take the form~\cite{Ask-arXiv2020}
\begin{subequations}
\begin{eqnarray}
\Phi_{\mathrm{R}}\left(x\right)&=&
e^{ikx}\Big[\theta\left(x_{1}-x\right)
+\sum_{m=1}^{3}t_{m}\theta\left(x-x_{m}\right)\theta\left(x_{m+1}-x\right)
\nonumber
\\
&&+t\theta\left(x-x_{4}\right)\Big],
\label{phiR}
\end{eqnarray}
\begin{eqnarray}
\Phi_{\mathrm{L}}\left(x\right)&=&
e^{-ikx}\Big[r\theta\left(x_{1}-x\right)
+\sum_{m=2}^{4}r_{m}\theta\left(x-x_{m-1}\right)\theta\left(x_{m}-x\right)\Big].
\nonumber
\\
\label{phiL}
\end{eqnarray}
\end{subequations}
Here $t_{m}$ ($r_{m}$) is the transmission (reflection) amplitude of the $m$th coupling point, $t$ ($r$) is the transmission (reflection) amplitude of the last (first) coupling
point, and $\theta\left(x\right)$ denotes the Heaviside step function. $x_{m}$ represents the $m$th (from left to right) coupling point of each configuration in Fig.~\ref{system}.

Substituting Eqs.~\eqref{phiR}-\eqref{phiL} into Eqs.~\eqref{motion1}-\eqref{motion3}, we can obtain
the expressions of the single-photon transmission and reflection amplitudes
\begin{widetext}
\begin{subequations}
\begin{equation}
t=\frac{-\left(\Delta_{a}-\Delta_{\mathrm{L},a}\right)\left(\Delta_{b}-\Delta_{\mathrm{L},b}\right)+\frac{1}{4}\left(\Gamma^{2}_{ab}-\Gamma_{a}\Gamma_{b}\right)+g_{ab}^{2}}{\left(i\Delta_{a}-i\Delta_{\mathrm{L},a}-\frac{1}{2}\Gamma_{a}\right)\left(i\Delta_{b}-i\Delta_{\mathrm{L},b}-\frac{1}{2}\Gamma_{b}\right)-\left(\frac{1}{2}\Gamma_{ab}+ig_{ab}\right)^{2}},
\label{transmission}
\end{equation}
\begin{equation}
r=\frac{\left\{\left[\frac{1}{2}\Gamma_{b}\left(i\Delta_{a}-i\Delta_{\mathrm{L},a}-\frac{1}{2}\Gamma_{a}\right)e^{i\frac{\alpha_{b}-\alpha_{a}}{2}}+\left(a\leftrightarrow b\right)\right]+\left(ig_{ab}+\frac{1}{2}\Gamma_{ab}\right)\sqrt{\Gamma_{a}\Gamma_{b}}\right\}e^{i\frac{\alpha_{a}+\alpha_{b}}{2}}}{\left(i\Delta_{a}-i\Delta_{\mathrm{L},a}-\frac{1}{2}\Gamma_{a}\right)\left(i\Delta_{b}-i\Delta_{\mathrm{L},b}-\frac{1}{2}\Gamma_{b}\right)-\left(\frac{1}{2}\Gamma_{ab}+ig_{ab}\right)^{2}},
\label{reflection}
\end{equation}
\end{subequations}
\end{widetext}
where $\Delta_{j}=\omega-\omega_{j}$ is the detuning between the photon and the atom $j$. 
One can see that the spectra are determined by some characteristic quantities~\cite{Kockum-PRL2018}, including the Lamb shifts $\Delta_{\mathrm{L},j}$, the individual decays $\Gamma_{j}$, the exchange interaction $g_{ab}$, the collective decay $\Gamma_{ab}$, and the phase factors $\alpha_{j}$, which can be defined as
\begin{subequations}
\begin{equation}	
\Delta_{\mathrm{L},j}=\sqrt{\gamma_{j1}\gamma_{j2}}\sin\left|\phi_{j2,j1}\right|,
\label{LS}
\end{equation}
\begin{equation}
\Gamma_{j}=\sum_{n,n'}\sqrt{\gamma_{jn}\gamma_{jn'}}\cos\phi_{jn,jn'},
\label{ID}
\end{equation}
\begin{equation}
g_{ab}=\frac{1}{2}\sum_{n,n'}\sqrt{\gamma_{an}\gamma_{bn'}}\sin\left|\phi_{bn,an'}\right|,
\label{EI}
\end{equation}
\begin{equation}
\Gamma_{ab}=\sum_{n,n'}\sqrt{\gamma_{an}\gamma_{bn'}}\cos\phi_{bn,an'},
\label{CD}
\end{equation}
\begin{equation}
\tan\alpha_{j}=\frac{\sum_{n,n'}\sqrt{\gamma_{jn}\gamma_{jn'}}\sin k\left(x_{jn}+x_{jn'}\right)}{\sum_{n,n'}\sqrt{\gamma_{jn}\gamma_{jn'}}\cos k\left(x_{jn}+x_{jn'}\right)},
\label{PhaseFactor}
\end{equation}
\end{subequations}
respectively, where $j=a,b$ and $n,n'=1,2$. $\gamma_{jn}=2V_{jn}^{2}/v_{\mathrm{g}}$ is the decay 
rate through the coupling point at $x_{jn}$. $\phi_{jn,j'n'}=\omega_{a}(x_{jn}-x_{j'n'})/v_{\mathrm{g}}$ 
is the phase acquired by the photon traveling from the connection point $x_{j'n'}$ to $x_{jn}$. In the 
above derivation,  we have assumed that the transition frequencies of the two atoms $\omega_{a}\approx\omega_{b}$, and made the Markov approximation. Thus  the wave vector $k$
in the definition of $\phi_{jn,j'n'}$ has been replaced by $\omega_{a}/v_{\mathrm{g}}$.
Note that Eqs.~\eqref{transmission} and \eqref{reflection} are the most general expressions 
for the scattering amplitudes, which are available for all the three types of topological configurations. 
One can further define the transmittance $T=|t|^{2}$ and the reflectance $R=|r|^{2}$. The conservation of photon number results in $T+R=1$.

Based on the general expressions of the scattering amplitudes given above, in the follows, we concentrate on two kinds of important phenomena, Fano interferences and EIT without control field,
in a wQED system containing double giant atoms.      
\section{\label{Fano Interferences}Fano interferences}
It is known that when a chain of two-level atoms strongly coupled to a waveguide, one can observe the emergence of asymmetric Fano lineshapes \cite{Tsoi-PRA2008,Cheng-OSA2012,Liao-PRA2015,Cheng-PRA2017,Mukhopadhyay-PRA2019} due to interference effects between the scattering amplitudes from different atoms. 
This is an example of the Fano interference phenomenon \cite{Fano-Phys.Rev1961,Miroshnichenko-Rev.Mod.Phys.2010}.
In addition, the appearance of Fano lineshapes is 
influenced by the relative phase picked up by the propagating photon when it travels from one atom to the next. Here, starting from the general expressions of the scattering amplitudes Eqs.~\eqref{transmission} and \eqref{reflection}, we investigate the similar phenomenon
in wQED systems with double giant atoms. Note that in addition to the phase delays, the Fano-like lineshapes in these systems are also dependent on the topological configurations. Thus they exhibit features different from the ones in wQED systems with two small atoms.      

In this section, for simplicity, we assume that the atoms have the same frequency with $\Delta_{j}=\Delta$, and all the bare decay rates are equal with $\gamma_{jn}=\gamma$. We also assume that the distances between neighboring points are equal with corresponding phase delay $\phi$. Without loss of generality, we set the phase corresponding to the leftmost point to be $0$ as reference. Under the above assumptions,
we discuss the spectral features, especially the conditions for the occurrence of Fano-like lineshapes, for the following three different topological configurations.

\subsection{Two separate giant atoms}
At first we consider the case of two separate giant atoms. From Eqs.~\eqref{LS}-\eqref{CD}, we can get the Lamb shifts $\Delta_{\mathrm{L},a}=\Delta_{\mathrm{L},b}=\Delta_{\mathrm{L}}=\gamma\sin\phi$, the individual decays $\Gamma_{a}=\Gamma_{b}=2\gamma\left(1+\cos\phi\right)$, the exchange interaction $g_{ab}=\gamma\left(\sin\phi+2\sin2\phi+\sin3\phi\right)/2$, and the collective decay $\Gamma_{ab}=
\gamma\left(\cos\phi+2\cos2\phi+\cos3\phi\right)$. 
Substituting these results into Eqs.~\eqref{transmission} and \eqref{reflection}, we can get the transmission and reflection amplitudes of this topology
\begin{subequations}
\begin{equation}
t=\frac{-\left(\Delta-\gamma\sin\phi\right)^{2}}{\left[i\Delta-\gamma\left(1+e^{i\phi}\right)\right]^{2}-\left[\frac{\gamma}{2}e^{i\phi}\left(1+e^{i\phi}\right)^{2}\right]^{2}},
\label{FT1}
\end{equation}
\begin{equation}
r=\frac{4i e^{3i\phi}\gamma\cos^{2}\left(\frac{\phi}{2}\right)\left[\Delta\cos 2\phi+\gamma\left(\sin\phi+\sin 2\phi\right)\right]}{\left[i\Delta-\gamma\left(1+e^{i\phi}\right)\right]^{2}-\left[\frac{\gamma}{2}e^{i\phi}\left(1+e^{i\phi}\right)^{2}\right]^{2}}.
\label{FR1}
\end{equation}
\end{subequations}
Without loss of generality, in the following part we focus on the reflectance $R=|r|^2$ only, for the transmittance $T$ and the reflectance $R$ are constrained by the relation $T+R=1$.
\begin{figure*}[t]
	\includegraphics[width=1\textwidth]{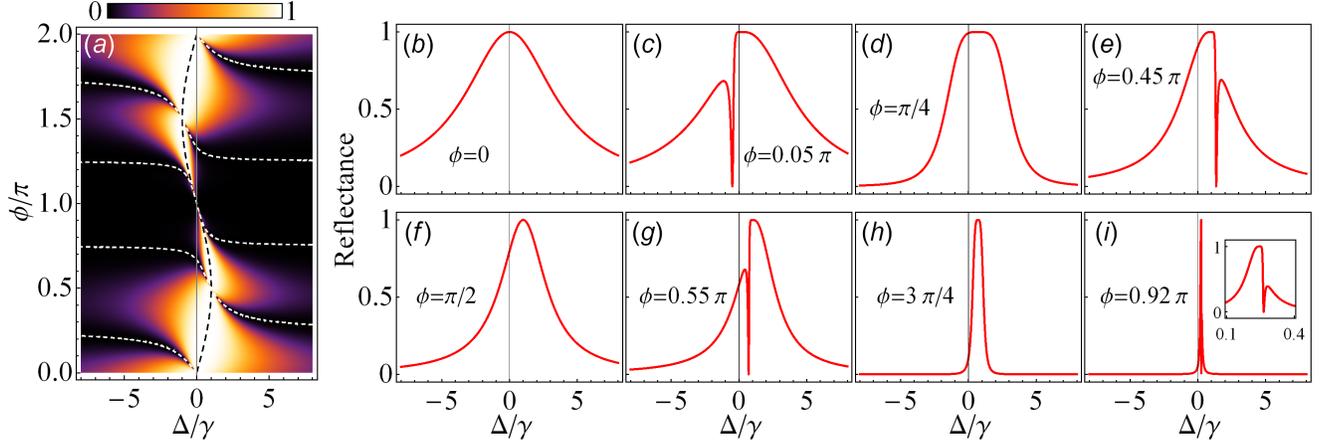}
	\caption{(a) Reflectance $R$ for two separate giant atoms as functions of detuning $\Delta$ and  phase $\phi$. The black dashed line is used to label the locations of the reflection peaks. The white dashed lines are used to label the reflection minima. The curves in (b)-(i) show cross sections of panel (a) at phases: (b) $\phi=0$, (c) $\phi=0.05\pi$, (d) $\phi=\pi/4$, (e) $\phi=0.45\pi$, (f) $\phi=\pi/2$, (g) $\phi=0.55\pi$, (h) $\phi=3\pi/4$, and (i) $\phi=0.92\pi$. }
	\label{F1}
\end{figure*}

In Fig.~\ref{F1}(a), we plot the reflectance $R$ as functions of $\Delta$ and $\phi$. 
Note that the spectra changes periodically with $\phi$, thus without loss of generality, the range of $\phi$ is chosen as a period $\phi\in[0,2\pi]$.
From Eq.~\eqref{FT1} and Eq.~\eqref{FR1}, we 
can obtain that the reflection peaks with $R=1$ appear at $\Delta=\Delta_{\mathrm{L}}=\gamma\sin\phi$ [labeled by the black dashed line in 
Fig.~\ref{F1}(a)]. And the reflection 
minima with $R=0$ appear at $\Delta=-\gamma\left(\sin\phi+\sin2\phi\right)/\cos2\phi$ [labeled by the white dashed line in Fig.~\ref{F1}(a)]. 
In addition, for some $\phi\in[0, \pi]$, we have relation $R(\Delta,\phi)=R(-\Delta,2\pi-\phi)$. Thus, without loss of generality, we show
in Figs.~\ref{F1}(b)-\ref{F1}(i) the cross sections at some typical phase delays in the region $\phi\in[0,\pi]$. 

Specifically, when $\phi=0$, $\phi=\pi/4$, $\phi=\pi/2$, and $\phi=3\pi/4$, the spectra are symmetric about $\Delta=\Delta_{\mathrm{L}}$.  In 
addition, when $\phi=0$ and $\phi=\pi/2$, the reflection spectra exhibit Lorentzian lineshapes [see Figs.~\ref{F1}(b) and \ref{F1}(f)]. And when $\phi=\pi/4$ and $\phi=3\pi/4$, the spectra exhibit super-Gaussian lineshapes [see Figs.~\ref{F1}(d) and \ref{F1}(h)]. Note that this
type of lineshape appears around the maximum reflection point when $\phi$ takes 
the values other than $0$ and $\pi/2$. In fact, the reflection coefficients  around the maximum reflection point $\Delta=\Delta_{\mathrm{L}}$ can be 
approximated as $R\approx1-\Delta'^{4}/(4g_{ab}^{4}+\Gamma_{ab}^{2}g_{ab}^{2})\approx\mathrm{exp}[-\Delta'^{4}/(4g_{ab}^{4}+\Gamma_{ab}^{2}g_{ab}^{2})]$, where $\Delta'=\Delta-\Delta_{\mathrm{L}}$. Therefore, the lineshapes exhibit super-Gaussian 
characteristic around the reflection peaks, as shown by Figs.~\ref{F1}(c)-\ref{F1}(e) and Figs.~\ref{F1}(g)-\ref{F1}(i). This is different from the Lorentzian lineshape displayed 
in Figs.~\ref{F1}(a) and \ref{F1}(f), which shows a Gaussian distribution near the reflection peak.

When $\phi$ takes values other than $\phi=0$, $\phi=\pi/4$, $\phi=\pi/2$, and $\phi=3\pi/4$, the spectra become asymmetric and  there appear reflection minima at $\Delta=-\gamma\left(\sin\phi+\sin2\phi\right)/\cos2\phi$, as shown in Figs.~\ref{F1}(c), \ref{F1}(e), \ref{F1}(g), and \ref{F1}(i).
Through the following analysis, one can find that in some regime,  the spectra around the reflection minima can be approximated as asymmetric Fano lineshapes. To this end, we rewrite the reflection amplitude Eq.~\eqref{FR1} as the superposition of two Lorentz spectra $r=r_{+}+r_{-}$, where
\begin{equation}
r_{\pm}=\frac{\pm e^{3i\phi}\Gamma_{\pm}}{i\left(\Delta-\Delta_{\pm}\right)-\Gamma_{\pm}},
\label{Fanoshape r1}
\end{equation}
with
\begin{subequations}
\begin{equation}
\Delta_{\pm}=\gamma\sin\phi\left(1\pm2\cos\phi\pm2\cos^{2}\phi\right),
\label{FanoDetuingS}
\end{equation}
\begin{equation}
\Gamma_{\pm}=\gamma\left(1+\cos\phi\right)\left(1\pm\cos2\phi\right)
\label{FanoDecayS}
\end{equation}
\end{subequations}
being the resonance points and the half widths, respectively.  To obtain an asymmetric Fano-type spectrum, the condition $\Gamma_{+}\gg\Gamma_{-}$ (or $\Gamma_{-}\gg\Gamma_{+}$) should be satisfied. Here the mode with large width denotes a continuum while the mode with small width represents a discrete level \cite{Cheng-OSA2012}. We can prove that under the condition $\Gamma_{\pm}\gg\Gamma_{\mp}$, the reflection coefficient around $\Delta_{\mp}$ can be approximated as
\begin{equation}
R\simeq\frac{\mathcal{F}\left(q+\epsilon\right)^{2}}{1+\epsilon^2},
\label{FanoLineshape}
\end{equation}
exhibiting a standard Fano lineshape \cite{Fano-Phys.Rev1961,Miroshnichenko-Rev.Mod.Phys.2010}. Here $q=(\Delta_{\pm}-\Delta_{\mp})/\Gamma_{\pm}$ is an asymmetry parameter, $\epsilon=(\Delta-\Delta_{\mp})/\Gamma_{\mp}$  is a reduced detuning, and $\mathcal{F}=\Gamma_{\pm}^{2}/[(\Delta_{\pm}-\Delta_{\mp})^{2}+\Gamma_{\pm}^{2}]$. 
Specifically, according to Eq.~\eqref{FanoDecayS}, we find that when the phase delay (restricted within half period) is taken as $\phi\in(0,0.1\pi)\cup(0.9\pi,\pi)$, the ratio $\Gamma_{+}/\Gamma_{-}>10$. Thus we can say that in this region  $\Gamma_{+}\gg\Gamma_{-}$ is satisfied, the reflection spectra for this case exhibits Fano lineshapes around the reflection minima $\epsilon=-q$, as shown in Figs.~\ref{F1}(c) and \ref{F1}(i). And when $\phi\in(0.4\pi,\pi/2)\cup(\pi/2,0.6\pi)$, we have $\Gamma_{-}/\Gamma_{+}>10$. Thus we can say that $\Gamma_{-}\gg\Gamma_{+}$ in this region. The corresponding reflection spectra containing Fano minima are shown in Figs.~\ref{F1}(e) and \ref{F1}(g).

Compared with the case of 
two small atoms \cite{Tsoi-PRA2008,Cheng-OSA2012,Liao-PRA2015,Cheng-PRA2017,Mukhopadhyay-PRA2019}, the shapes of the
spectra for the case of two separate giant atoms (shown in Fig.~\ref{F1}) are similar. The main difference is that for two small atoms, the 
reflection maxima are always located at $\Delta=0$ \cite{Mukhopadhyay-PRA2019}, but for two separate giant atoms, 
the reflection maxima  appear at $\Delta=\gamma\sin\phi$. The reason is that two separate giant atoms in a wQED structure can be
looked on as two small atoms with frequency shift $\gamma\sin\phi$, effective decay $2\gamma\left(1+\cos\phi\right)$, and phase delay 
$2\phi$ (proportional to effective distance between the atoms). However, for those unique configurations of the multi-giant-atom 
systems, i.e., the topologies with two braided or nested giant atoms, there is no similar analogy. Thus the spectra become more different. We will discuss these cases in detail in the following subsections.
\subsection{Two braided giant atoms}
\begin{figure}[t]
	\includegraphics[width=0.5\textwidth]{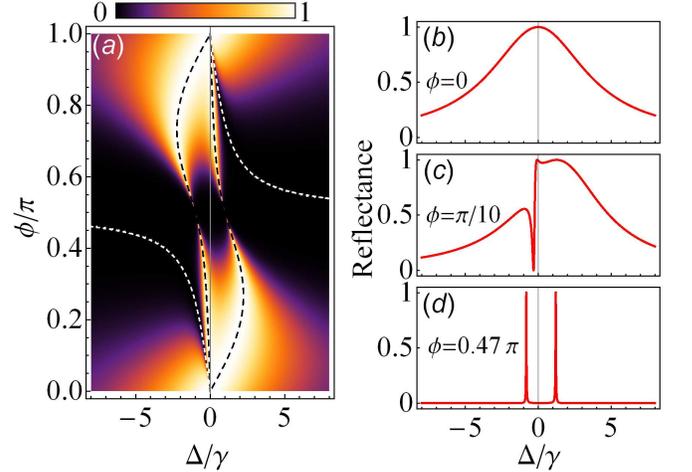}
	\caption{(a) Reflectance $R$ for two braided giant atoms as functions of detuning $\Delta$ and  phase $\phi$. The black dashed line is used to label the locations of the reflection peaks. The white dashed lines are used to label the reflection minima. The curves in (b)-(d) show cross sections of panel (a) at phases: (b) $\phi=0$, (c) $\phi=\pi/10$, and (d) $\phi=0.47\pi$.
		 }
	\label{F2}
\end{figure}

Now we consider the case of two braided giant atoms. From Eqs.\eqref{LS}-\eqref{CD}, we can get the Lamb shifts $\Delta_{\mathrm{L},a}=\Delta_{\mathrm{L},b}=\Delta_{\mathrm{L}}=\gamma\sin2\phi$, the individual decays $\Gamma_{a}=\Gamma_{b}=2\gamma\left(1+\cos2\phi\right)$, the exchange interaction $g_{ab}=\gamma\left(3\sin\phi+\sin3\phi\right)/2$, and the collective decay $\Gamma_{ab}=
\gamma\left(3\cos\phi+\cos3\phi\right)$. 
Thus the transmission and reflection amplitudes of this configuration can be written as
\begin{subequations}
\begin{equation}
t=\frac{-\left(\Delta-\gamma\sin 2\phi\right)^{2}+\gamma^{2}\left(\sin^{2}2\phi+\sin^{2}\phi\right)}{\left[i\Delta-\gamma\left(1+e^{2i\phi}\right)\right]^{2}-\left[\frac{\gamma}{2}\left(3e^{i\phi}+e^{3i\phi}\right)\right]^{2}},
\label{FT2}
\end{equation}
\begin{equation}
r=\frac{4ie^{3i\phi}\gamma\cos^{2}\phi\left(\Delta\cos\phi+\gamma\sin\phi\right)}{\left[i\Delta-\gamma\left(1+e^{2i\phi}\right)\right]^{2}-\left[\frac{\gamma}{2}\left(3e^{i\phi}+e^{3i\phi}\right)\right]^{2}}.
\label{FR2}
\end{equation}
\end{subequations}
We plot the reflectance $R$ as functions of $\Delta$ and $\phi$ [see Fig.~\ref{F2}(a)]. One can find that the period of the spectra is $\pi$.  
Different from the case of two separate atoms, there are
two reflection peaks appearing at $\Delta=\gamma\sin 2\phi\pm\gamma\sqrt{1-\cos\phi\cos3\phi}$, except for  some special phases $\phi=n\pi$ 
, as illustrated by the black dashed lines.  
And the reflection minima with $R=0$ appear at $\Delta=-\gamma\tan\phi$, as shown by the white dashed lines.  Moreover, the two reflection 
peaks are always located at the same side of the minimum point [right when $\phi\in(0,\pi/2)$, and left when $\phi\in(\pi/2,\pi)$].
Finally, for some $\phi\in[0, \pi/2]$, we have relation $R(\Delta,\phi)=R(-\Delta,\pi-\phi)$. Thus, without loss of generality, we show
in Figs.~\ref{F2}(b)-\ref{F2}(d) the cross sections at some typical phase delays in the region $\phi\in[0,\pi/2]$. 

Specifically, when $\phi=0$, the reflection spectrum has a Lorentzian lineshape centred at $\Delta=0$ with width $8 \gamma$, as shown
in Fig.~\ref{F2}(b). With $\phi$ increasing, the reflection peak splits into two, and a reflection minimum appears at $\Delta=-\gamma\tan\phi$, 
as shown in Fig.~\ref{F2}(c) (with $\phi=\pi/10$). Similar to the case of two separate giant atoms, we can prove that if the phase $\phi$ is appropriately chosen, the spectrum near the reflection minimum exhibits a Fano lineshape. Again, to analyze the mechanism of Fano interference, we rewrite the reflection amplitude Eq.~\eqref{FR2} as the sum of two Lorentz-type amplitudes $r=r_{+}+r_{-}$, where $r_{\pm}$ can also be expressed in terms of  Eq.~\eqref{Fanoshape r1}, with
\begin{subequations}
\begin{equation}
\Delta_{\pm}=\gamma\left(\sin2\phi\pm\frac{3}{2}\sin\phi\pm\frac{1}{2}\sin3\phi\right),
\label{FanoDetuningB}
\end{equation}
\begin{equation}
\Gamma_{\pm}=\gamma\left(1+\cos2\phi\right)\left(1\pm\cos\phi\right).
\label{FanoDecayB}
\end{equation}
\end{subequations}
And straightforwardly, under the condition $\Gamma_{\pm}\gg\Gamma_{\mp}$, the reflection spectrum around $\Delta_{\mp}$ can be fitted by a Fano lineshape described by Eq.~\eqref{FanoLineshape}. From Eq.~\eqref {FanoDecayB}, we can see that to obtain Fano lineshapes, the phase delay (restricted within one period) should be taken as  $\phi\in(0,0.19\pi)$ to ensure
$\Gamma_{+}\gg\Gamma_{-}$, or $\phi\in(0.81\pi,\pi)$ to ensure $\Gamma_{-}\gg\Gamma_{+}$. Thus the reflection spectrum shown in Figs.~\ref{F2}(c) (with $\phi=\pi/10$, leading to $\Gamma_{+}\gg\Gamma_{-}$), exhibits a Fano lineshapes near the reflection minimum. 

When $\phi$ is closed to $\pi/2$ (not equal), the spectra exhibit lineshapes like 
vacuum Rabi splitting, as shown in Fig.~\ref{F2}(d). This kind of spectroscopic characteristics can be explained in terms of 
a \textit{nearly} decoherence-free interaction between two braided giant atoms.
Note that when $\phi=\pi/2$, the atoms have vanished individual decays $\Gamma_{a}
=\Gamma_{b}=0$ and meanwhile preserve non-zero exchange interaction  $g_{ab}=\gamma$ between them,
called decoherence-free interaction \cite{Kockum-PRL2018}. However, if 
the phase $\phi$ is exactly equal to $\pi/2$, this phenomenon can not be probed by the photon scattering spectra for the atoms are decoupled 
from the waveguide. Thus, in Fig.~\ref{F2}(d), we let  the  phase $\phi$ slightly deviate from $\pi/2$, which can ensure that
$g_{ab}\simeq\gamma$ and at the same time $\Gamma_i$ ($i=a,b$) obtains a small value satisfying  $\Gamma_i\ll g_{ab}$.
In this regime, the incident photon can interact with the system, therefore the nearly decoherence-free interaction
can be probed. Moreover, in this case the system works in the strong coupling regime with the exchange interaction being 
much larger than the
individual decays of atoms. Thus it is not surprising that  we can obtain
a vacuum Rabi splitting like spectrum in Fig.~\ref{F2}(d).    
To see this more clearly, we set $\phi={\pi}/{2}+\delta$ (with $\left|\delta\right|\ll1$), and write down the corresponding approximate expressions
of the scattering amplitudes
\begin{subequations}
	\begin{equation}
	t\approx\frac{-\left(\Delta+2\gamma\delta\right)^{2}+\gamma^{2}}{i\left(\Delta+2\gamma\delta\right)\left[i\left(\Delta+2\gamma\delta\right)-4\gamma\delta^{2}\right]+\gamma^{2}},
	\label{FT31}
	\end{equation}
	\begin{equation}
	r\approx\frac{4\gamma^{2}\delta^{2}}{i\left(\Delta+2\gamma\delta\right)\left[i\left(\Delta+2\gamma\delta\right)-4\gamma\delta^{2}\right]+\gamma^{2}}.
	\label{FR31}
	\end{equation}
\end{subequations}
Clearly, these expressions represent standard vacuum Rabi splitting spectra with two peaks located at 
$\Delta=-2\gamma\delta\pm\gamma$. The distance between the two peaks is $2g_{ab}=2\gamma$, 
characterizing the strength of the exchange interaction between atoms. The width of each peak is 
$(\Gamma_a+\Gamma_b)/2=4\gamma\delta^{2}$. These results are in accordance with the spectrum 
shown in Fig.~\ref{F2}(d), where the phase is taken as $\phi=0.47\pi$. In summary, the spectra near 
$\phi=\pi/2$ can be used to sensitively probe the decoherence-free interaction between two braided giant 
atoms. Moreover, this result indicates that the system in this regime can be regarded as an effective cavity QED structure, and may
have potential applications in quantum information processings. 

\begin{figure*}[t]
	\includegraphics[width=1\textwidth]{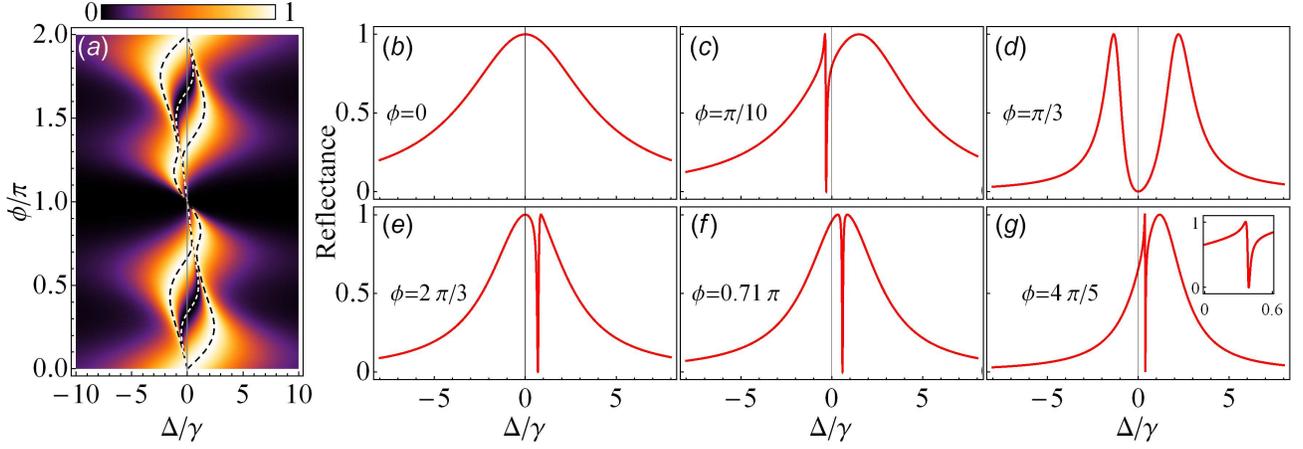}
	\caption{(a) Reflectance $R$ for two nested giant atoms as functions of detuning $\Delta$ and  phase $\phi$. The Black dashed line is used to label the locations of the reflection peaks. The white dashed lines are used to label the reflection minima. The curves in (b)-(g) show cross sections of panel (a) at phases: (b) $\phi=0$, (c) $\phi=\pi/10$, (d) $\phi=\pi/3$, (e) $\phi=2\pi/3$, (f) $\phi=0.71\pi$, (g) $\phi=4\pi/5$. 	}
	\label{F3}
\end{figure*}
\subsection{Two nested giant atoms}
Finally, we consider the case of two nested giant atoms. From Eqs.~\eqref{LS}-\eqref{CD}, we can get the 
Lamb shifts $\Delta_{\mathrm{L},a}=\gamma\sin3\phi$, $\Delta_{\mathrm{L},b}=\gamma\sin\phi$, the individual decays $\Gamma_{a}=2\gamma\left(1+\cos3\phi\right)$, $\Gamma_{b}=2\gamma\left(1+\cos\phi\right)$, the exchange interaction $g_{ab}=\gamma\left(\sin\phi+\sin2\phi\right)$, and the collective decay $\Gamma_{ab}=
2\gamma\left(\cos\phi+\cos2\phi\right)$.
Thus the transmission and reflection amplitudes of this topology can be written as
\begin{subequations}
\begin{equation}
t=\frac{-\left(\Delta-\gamma\sin3\phi\right)\left(\Delta-\gamma\sin\phi\right)+\gamma^{2}\left(\sin\phi+\sin 2\phi\right)^{2}}{[i\Delta-\gamma(1+e^{3i\phi})][i\Delta-\gamma(1+e^{i\phi})]-[\gamma e^{i\phi}(1+e^{i\phi})]^{2}},
\label{FT3}
\end{equation}
\begin{equation}
r=\frac{4ie^{3i\phi}\gamma\cos^{2}\frac{\phi}{2}[\Delta(2-2\cos\phi+\cos 2\phi)-\gamma(\sin\phi-\sin 2\phi)]}{[i\Delta-\gamma(1+e^{3i\phi})][i\Delta-\gamma(1+e^{i\phi})]-[\gamma e^{i\phi}(1+e^{i\phi})]^{2}}.
\label{FR3}
\end{equation}
\end{subequations}
We plot the reflectance $R$ as function of $\Delta$ and $\phi$ [see Fig.~\ref{F3}(a)]. The period 
of the spectra is $2\pi$. The black dashed lines are used to label the location of reflection peaks at 
\begin{eqnarray}
&&\Delta=\frac{1}{2}\gamma(\sin 3\phi+\sin\phi)
\nonumber
\\
&&\pm\gamma\sqrt{(\sin\phi+\sin 2\phi)^{2}+\frac{1}{4}(\sin 3\phi-\sin\phi)^2},
\nonumber
\end{eqnarray}
 while the white dashed line represents the reflection minima at $\Delta=\gamma(\sin\phi-\sin 2\phi)/(2-2\cos\phi+\cos 2\phi)$, 
except for some special phases $\phi=n\pi$. Note that for some $\phi\in[0, \pi]$, we have relation 
$R(\Delta,\phi)=R(-\Delta,2\pi-\phi)$. Thus, for simplicity, we only consider the cross sections at some 
typical phase delays in the region $\phi\in[0,\pi]$, as shown in Figs.~\ref{F3}(b)-\ref{F3}(g).

Specifically, when $\phi=0$, the reflection spectrum exhibits a Lorentzian lineshape centered at 
$\Delta=0$ with width $8\gamma$, as shown in Fig.~\ref{F3}(b). When $0<\phi<\pi$, a reflection minimum
with $R=0$ appears between two reflection peaks, as shown in Figs.~\ref{F3}(c)-\ref{F3}(g). 
Note that this is different from the braided atoms, where both reflection peaks are located at the same side of the reflection minimum.  The distribution of the reflection peaks and dips 
for different phases is summarized in more detail in what follows. When $\phi<\pi/3$, both the left peak and 
the reflection minimum appear at $\Delta<0$, and the right peak is located at $\Delta>0$ [see Fig.~\ref{F3}(c)]. When $\phi=\pi/3$, the reflection 
minimum appears at $\Delta=0$ [see Fig.~\ref{F3}(d)]. When $\pi/3<\phi<2\pi/3$, the left peak  is located 
at $\Delta<0$, and the reflection minimum and the right peak appear at $\Delta>0$. When $\phi=2\pi/3$, the left reflection peak  is located 
at $\Delta=0$, and the reflection minimum and the right peak appear at $\Delta>0$ [see Fig.~\ref{F3}(e)]. When $\phi>2\pi/3$, all the reflection peaks and the reflection minimum appear at $\Delta>0$ [see Fig.~\ref{F3}(f) and \ref{F3}(g)], and particularly, when $\phi\simeq0.71\pi$, the spectrum becomes symmetric [see Fig.~\ref{F3}(f)].

Similar to the other two configurations, if the phase $\phi$ is appropriately chosen, the spectrum near the reflection minimum exhibits a Fano 
lineshape. Also, to better understand the Fano interference, we decompose the reflection amplitude Eq.~\eqref{FR3} into the sum of two 
terms $r=r_{+}+r_{-}$, where
\begin{equation}
r_{\pm}=\frac{\chi_{\pm}\Gamma_{\pm}}{i\left(\Delta-\Delta_{\pm}\right)-\Gamma_{\pm}}
\label{Fanoshape r3}
\end{equation}
are Lorentz-type amplitudes with 
\begin{subequations}
\begin{equation}
\Delta_{\pm}=\gamma\left[\frac{1}{2}\left(\sin\phi+\sin3\phi\right)\mp\sqrt{2A}\left|\cos\frac{\phi}{2}\right|\cos\left(2\phi+\zeta\right)\right],
\label{FanoDetuingN}
\end{equation}
\begin{equation}
\Gamma_{\pm}=\gamma\left[1+\frac{1}{2}\cos\phi+\frac{1}{2}\cos3\phi\pm\sqrt{2A}\left|\cos\frac{\phi}{2}\right|\sin\left(2\phi+\zeta\right)\right].
\label{FanoDecayN}
\end{equation}
\end{subequations}
The expressions of coefficients $\chi_{\pm}=\chi e^{i(\phi-\zeta\pm\vartheta)}$ and $A$ are given in Appendix \ref{DefinitionC}.
Again, one can prove that under the condition $\Gamma_{\pm}\gg\Gamma_{\mp}$, the reflection spectra around $\Delta_{\mp}$ can be fitted by a Fano lineshapes described by Eq.~\eqref{FanoLineshape}, where $q=\cos 2\vartheta(\Delta_{\mp}-\Delta_{\pm})/\Gamma_{\pm}\pm\sin 2\vartheta$, $\epsilon=(\Delta-\Delta_{\mp})/\Gamma_{\mp}$, and $\mathcal{F}=\chi^{2}\Gamma_{\pm}^{2}/[(\Delta_{\pm}-\Delta_{\mp})^{2}+\Gamma_{\pm}^{2}]$. After some calculations, we can find that when the phase delay (restricted within half period) is chosen as $\phi\in (0, 0.23\pi)$, the ratio $\Gamma_{+}/\Gamma_{-}>10$. Thus in  this regime  $\Gamma_{+}\gg\Gamma_{-}$ is satisfied. The interference between modes $r_{+}$ and  $r_{-}$ results in a Fano lineshape around the reflection minimum [see Figs.~\ref{F3}(c)]. And when 
 $\phi\in (0.51\pi, \pi)$, we have $\Gamma_{-}/\Gamma_{+}>10$. Thus $\Gamma_{-}\gg\Gamma_{+}$ is satisfied and the Fano lineshape also appears [see Figs.~\ref{F3}(e)-(g)]. In particular, when $\phi\simeq0.71\pi$, the asymmetric factor $q=0$, the lineshape
 becomes symmetric about $\Delta\simeq0.59\gamma$ [see Figs.~\ref{F3}(f)]. 

It should be pointed out that the spectra in Figs.~\ref{F3}(c)-\ref{F3}(g) look like EIT or  Autler-Townes splitting (ATS) spectra, but according to
our analysis in the next section, only the curve in Fig.~\ref{F3}(d) exhibits an asymmetric ATS spectrum with a reflection minimum at $\Delta=0$. The curves in the other figures are neither EIT nor ATS spectra.  In the next section, we will provide general conditions to generate EIT or ATS spectra in the systems consisting of two nested giant atoms.

\section{\label{EITwithoutCF}Single-photon EIT without control field}
The  control-field-free EIT phenomenon was firstly investigated in the wQED system  
containing double small atoms \cite{Shen-PRB2007,Fang-PRA2017,Mukhopadhyay-PRA2020}. Similar phenomenon in the double giant-atom 
systems for some special cases, e.g., two 
braided giant atoms with the phase delay between neighboring coupling points being set as 
$\pi$, was also investigated \cite{Ask-arXiv2020}. Here we aim to provide the general conditions to produce EIT without control field in wQED systems containing double giant atoms. To this end, 
in what follows we will firstly 
derive the conditions for EIT through analyzing the master equation of system, and then obtain the 
expressions of the scattering amplitudes under these conditions, based on the 
general results provided in Sec.~\ref{GeneralSolutions}. We find that compared with small atoms, giant atoms possess more working points to realize EIT-type phenomenon. For example, for two small atoms, to obtain EIT, detuning between the atoms is required, which plays the role of control field \cite{Shen-PRB2007,Fang-PRA2017,Mukhopadhyay-PRA2020}. But for double giant atoms, there are
more choices. Even if the atoms are with equal frequencies, the effective control field can also be obtained by adjusting the Lamb shifts (through changing the phase delays between coupling points), which will be discussed in Sec.~\ref{S-A}.

By treating the incident single photon as a weak driving field, and in a frame rotating with the drive frequency $\omega$, the master equation for double-giant-atom wQED structures can be written as \cite{Kockum-PRL2018} 
\begin{eqnarray}
\dot{\hat{\rho}}&=&
-i\left[{\hat{H}_{\mathrm{drive}},\hat{\rho}}\right]+\sum_{j}\Gamma_{j}\mathcal{D}\left[\hat{\sigma}_{j}^{-}\right]\hat{\rho}
\nonumber \\
&&+\Gamma_{ab}\sum_{j\neq j'}\left(\hat{\sigma}_{j}^{-}\hat{\rho}\hat{\sigma}_{j'}^{+}-\frac{1}{2}\{\hat{\sigma}_{j}^{+}\hat{\sigma}_{j'}^{-},\hat{\rho}\}\right),
\label{master eq}
\end{eqnarray}
with
\begin{eqnarray}
\hat{H}_{\mathrm{drive}}&=&-\sum_{j}\left(\Delta_{j}-\Delta_{\mathrm{L},j}\right)\hat{\sigma}_{j}^{+}\hat{\sigma}_{j}^{-}+g_{ab}\left(\hat{\sigma}_{a}^{+}\hat{\sigma}_{b}^{-}+\hat{\sigma}_{b}^{+}\hat{\sigma}_{a}^{-}\right)
\nonumber \\
&&-\frac{i}{2}\sum_{j}\left(\Omega_{j}\hat{\sigma}_{j}^{+}-\mathrm{H.c.}\right),
\label{Hdrive}
\end{eqnarray}
where $\mathcal{D}[\hat{O}]\hat{\rho}=\hat{O}\hat{\rho}\hat{O}^{\dag}-\{\hat{O}^{\dag}\hat{O},\hat{\rho}\}/2$ is the Lindblad operator. The Rabi frequency of the atom $j$ is defined as $\Omega_{j}=\Omega_{j1}e^{i\phi_{j1,a1}}+\Omega_{j2}e^{i\phi_{j2,a1}}$, where $\Omega_{jn}=\sqrt{2\gamma_{jn}}\alpha$ is the Rabi frequency
at coupling points $x_{jn}$, and $|\alpha|^{2}$ is the number of photons per second coming from the coherent drive. The
other qualities  are the same as those defined in defined in Sec.~\ref{GeneralSolutions}.    
 
The key to obtain the EIT without control field in wQED system with giant atoms is to
generate dark (decoupled from the waveguide) and bright (coupled to the waveguide) modes and at the same time persist waveguide induced interactions between them. Specifically, the bright and the dark states can be either collective or single-atom states, and we will discuss these two cases in the following subsections. 
\subsection{\label{S-A}Realizing EIT using waveguide-mediated interactions between atomic collective states}
By writing the master equation \eqref{master eq} and the Hamiltonian \eqref{Hdrive} in the 
symmetric-antisymmetric (S-A) basis $\hat{\sigma}_{\mathrm{S,A}}^{-}=\left(\hat{\sigma}_{a}^{-}\pm\hat{\sigma}_{b}^{-}\right)/\sqrt{2}$,
we have
\begin{eqnarray}
\dot{\hat{\rho}}&=&
-i\left[{\hat{H}_{\mathrm{drive}},\hat{\rho}}\right]+\sum_{u}\Gamma_{u}\mathcal{D}\left[\hat{\sigma}_{u}^{-}\right]\hat{\rho}
\nonumber \\
&&+\Gamma_{\mathrm{SA}}\sum_{u\neq v}\left[\hat{\sigma}_{u}^{-}\hat{\rho}\hat{\sigma}_{v}^{+}-\frac{1}{2}\left(\hat{\sigma}_{u}^{+}\hat{\sigma}_{v}^{-}\hat{\rho}+\hat{\rho}\hat{\sigma}_{u}^{+}\hat{\sigma}_{v}^{-}\right)\right],
\label{master-eq-SA}
\end{eqnarray}
with
\begin{eqnarray}
\hat{H}_{\mathrm{drive}}&=&
-\sum_{u}\Delta_{u}\hat{\sigma}_{u}^{+}\hat{\sigma}_{u}^{-}+g_{\mathrm{SA}}\left(\hat{\sigma}_{\mathrm{S}}^{+}\hat{\sigma}_{\mathrm{A}}^{-}+\hat{\sigma}_{\mathrm{A}}^{+}\hat{\sigma}_{\mathrm{S}}^{-}\right)
\nonumber \\
&&-\frac{i}{2}\sum_{u}\left(\Omega_{u}\hat{\sigma}_{u}^{+}-\mathrm{H.c.}\right).
\label{Hamiltonian-SA}
\end{eqnarray}
Here, $u,v=\mathrm{S},\mathrm{A}$. The exchange interaction, the individual decays and 
the collective decay in the S-A basis are defined as
\begin{subequations}
\begin{equation}
g_{\mathrm{SA}}=-\frac{1}{2}\left(\Delta_{a}-\Delta_{\mathrm{L},a}-\Delta_{b}+\Delta_{\mathrm{L},b}\right)
\label{gSA}
\end{equation}
\begin{equation}
\Gamma_{\mathrm{S}}=\frac{1}{2}\left(\Gamma_{a}+\Gamma_{b}\right)+\Gamma_{ab},
\label{Gamma S}
\end{equation}
\begin{equation}
\Gamma_{\mathrm{A}}=\frac{1}{2}\left(\Gamma_{a}+\Gamma_{b}\right)-\Gamma_{ab},
\label{Gamma A}
\end{equation}
\begin{equation}
\Gamma_{\mathrm{SA}}=\frac{1}{2}\left(\Gamma_{a}-\Gamma_{b}\right).
\label{Gamma SA}
\end{equation}
\end{subequations}
The effective detuning and the Rabi frequency of the symmetric (or antisymmetric) mode are defined as $\Delta_{\mathrm{S,A}}=\sum_{j=a,b}(\Delta_{j}-\Delta_{\mathrm{L},j})/2\mp g_{ab}$ and $\Omega_{\mathrm{S,A}}=(\Omega_{a}\pm\Omega_{b})/\sqrt{2}$, respectively.

To achieve master equation that can describe EIT or ATS type dynamics, the following 
conditions should be satisfied: (i) One of the collective modes is coupled to the waveguide, forming a bright state, and the 
other should be decoupled from the waveguide, forming a dark state; (ii) The collective decay should be zero;  (iii) The exchange interaction between the symmetric and the antisymmetric modes should be non-zero and plays the role of a control field.  For example, we can 
choose the symmetric state $|\mathrm{S}\rangle=\hat{\sigma}_{\mathrm{S}}^{+}|gg\rangle$ as a dark state, and the antisymmetric state $|\mathrm{A}\rangle=\hat{\sigma}_{\mathrm{A}}^{+}|gg\rangle$ as a bright state, i.e., $\Gamma_{\mathrm{S}}=0$, $\Gamma_{\mathrm{A}}\neq0$, $\Gamma_{\mathrm{SA}}=0$, and $g_ {\mathrm{SA}}\neq 0$, 
the master equation \eqref{master-eq-SA} and the Hamiltonian \eqref{Hamiltonian-SA} become
\begin{eqnarray}
\dot{\hat{\rho}}&=&-i\left[{\hat{H},\hat{\rho}}\right]+\Gamma_{A}\mathcal{D}\left[\hat{\sigma}_{A}^{-}\right]\hat{\rho}
\label{master-eq-EIT-driveA}
\end{eqnarray}
and
\begin{eqnarray}
\hat{H}&=&
-\sum_{u}\Delta_{u}\hat{\sigma}_{u}^{+}\hat{\sigma}_{u}^{-}+g_{\mathrm{SA}}\left(\hat{\sigma}_{\mathrm{S}}^{+}\hat{\sigma}_{\mathrm{A}}^{-}+\hat{\sigma}_{\mathrm{A}}^{+}\hat{\sigma}_{\mathrm{S}}^{-}\right)
\nonumber \\
&&-\frac{i}{2}\left(\Omega_{\mathrm{A}}\hat{\sigma}_{\mathrm{A}}^{+}-\mathrm{H.c.}\right),
\label{Hamiltonian-EIT-driveA}
\end{eqnarray}
respectively. We can prove that Eqs.~\eqref {master-eq-EIT-driveA} and \eqref{Hamiltonian-EIT-driveA}  can be mapped to the equation of motion describing 
the dynamics of a driven $\Lambda$-type atom \cite{Fleischhauer-Rev.Mod.Phys.2005} that can 
generate EIT or ATS type scattering spectra. The role of the control field is played by the exchange interaction $g_{\mathrm{SA}}$. 
Specifically, we consider a three-level $\Lambda$-type atom with a ground sate $|0\rangle$, a metastable state  $|1\rangle$, and an excited  state  $|2\rangle$, the transition $|0\rangle  \leftrightarrow |2\rangle$ ($|1\rangle  \leftrightarrow |2\rangle$) is coupled by a probe (control) field with Rabi frequency $\Omega_{\mathrm{p}}$ ($\Omega_{\mathrm{c}}$) to generate EIT phenomenon. After analyzing the master equation of this system (see Appendix~\ref{mapping-to-Lambda}),  we can find the following  analogies: $|gg\rangle\leftrightarrow|0\rangle$, $|\mathrm{S}\rangle\leftrightarrow|1\rangle$, $|\mathrm{A}\rangle\leftrightarrow|2\rangle$, $\hat{\sigma}_{\mathrm{S}}^{-}\leftrightarrow\hat{\sigma}_{01}$, $\hat{\sigma}_{\mathrm{A}}^{-}\leftrightarrow\hat{\sigma}_{02}$, $g_{\mathrm{SA}}\leftrightarrow\Omega_{\mathrm{c}}/2$, $\Omega_{\mathrm{A}}\leftrightarrow\Omega_{\mathrm{p}}$, $\Delta_{\mathrm{S}}\leftrightarrow\Delta_{\mathrm{p}}-\Delta_{\mathrm{c}}$, $\Delta_{\mathrm{A}}\leftrightarrow\Delta_{\mathrm{p}}$, $\Gamma_{\mathrm{A}}\leftrightarrow\Gamma_{20}$. 

To verify above analysis, we simplify Eqs.~\eqref{transmission}-\eqref{reflection} under the conditions $\Gamma_{\mathrm{S}}=0$, $\Gamma_{\mathrm{A}}\neq0$, $\Gamma_{\mathrm{SA}}=0$, and $g_ {\mathrm{SA}}\neq 0$, and obtain the following expressions of transmission and reflection amplitudes 
\begin{subequations}
\begin{equation}
t=\frac{-\Delta_{\mathrm{S}}\Delta_{\mathrm{A}}+g_{\mathrm{SA}}^{2}}{i\Delta_{\mathrm{S}}\left(i\Delta_{\mathrm{A}}-\frac{\Gamma_{\mathrm{A}}}{2}\right)+g_{\mathrm{SA}}^{2}},
\label{transmissionEIT-Amode}
\end{equation}
\begin{equation}
r=\frac{\frac{1}{2}i\Gamma_{\mathrm{A}}\Delta_{\mathrm{S}}}{i\Delta_{\mathrm{S}}\left(i\Delta_{\mathrm{A}}-\frac{\Gamma_{\mathrm{A}}}{2}\right)+g_{\mathrm{SA}}^{2}},
\label{reflectionEIT-Amode}
\end{equation}
\end{subequations}
which represent well known EIT or ATS spectra, depending on the strength of ``control field" $g_{\mathrm{SA}}$.  When $\Delta_{\mathrm{S}}\approx\Delta_{\mathrm{A}}$, we replace them with $\mathcal{Z}$, and inspect
the complex roots of the denominator of the scattering amplitude \cite{Abi-Salloum-Phys.Rev.A2010,Anisimov-Phys.Rev.Lett.2011}, 
\begin{equation}
\mathcal{Z}_{\pm}=-i\frac{\Gamma_{\mathrm{A}}}{4}\pm\frac{1}{4}\sqrt{16g_{\mathrm{SA}}^{2}-\Gamma_{\mathrm{A}}^{2}},
\end{equation}
which are purely imaginary for
\begin{equation}
|g_{\mathrm{SA}}|<\frac{\Gamma_{\mathrm{A}}}{4}.
\label{EITregimeSA-driveA}
\end{equation}
In this parameter regime, the transmission point located at $\Delta_{\mathrm{S}}=0$ is caused by destructive interference between two resonances. This regime is so called EIT regime. Otherwise, the system enters the ATS regime, in which the scattering
spectrum is made up of two peaks corresponding to the dressed states, and  the observed dip can be interpreted as a gap between the two peaks.

Alternatively, one can also choose the state $|\mathrm{A}\rangle$ as a dark state, and the state $|\mathrm{S}\rangle$ as a bright state, i.e., 
$\Gamma_{\mathrm{A}}=0$, $\Gamma_{\mathrm{S}}\neq0$, $\Gamma_{\mathrm{SA}}=0$, and 
$g_{\mathrm{SA}}\neq 0$, to achieve EIT phenomenon. All the results for this case can be obtained from 
Eqs.~\eqref {master-eq-EIT-driveA}-\eqref{EITregimeSA-driveA} by the index exchange
$\mathrm{S}\leftrightarrow\mathrm{A}$. 

As specific examples of above general results, in the follows we consider the maximum symmetric case, with 
equal bare decay rates $\gamma_{jn}=\gamma$, and equal phase delay $\phi$ between neighboring points. Without loss of generality, we set the phase corresponding to the leftmost coupling point is zero. In Figs.~\ref{E1}(a)-\ref{E1}(c), we plot $\Gamma_{\mathrm{S}}$, $\Gamma_{\mathrm{A}}$ and $\Gamma_{\mathrm{SA}}$ as the functions of $\phi$ for three different topologies. We use the blue arrows to indicate the points 
where EIT occurs with $\Gamma_{\mathrm{S}}=0$, $\Gamma_{\mathrm{A}}\neq0$ (or $\Gamma_{\mathrm{S}}\neq 0$, $\Gamma_{\mathrm{A}}=0$) and $\Gamma_{\mathrm{SA}}=0$. The corresponding reflection spectra are shown in the insets.  In the follows we will discuss this issue in details for different configurations. 

\subsubsection{Two separate giant atoms}
For two separate giant atoms, from Eqs.~\eqref{gSA}-\eqref{Gamma SA}, we can derive the exchange interaction and effective decay rates for the maximum symmetric case
\begin{subequations}
\begin{equation}
g_{\mathrm{SA}}=\frac{1}{2}\Delta_{ab},
\label{gSA1}
\end{equation}
\begin{equation}
\Gamma_{\mathrm{A}}=\gamma\left(2+\cos\phi-2\cos 2\phi-\cos 3\phi\right),
\label{GammaA1}
\end{equation}
\begin{equation}
\Gamma_{\mathrm{S}}=\gamma\left(2+3\cos\phi+2\cos 2\phi+\cos 3\phi\right),
\label{GammaS1}
\end{equation}
\begin{equation}
\Gamma_{\mathrm{SA}}=0.
\label{GammaSA1}
\end{equation}
\end{subequations}
Here, $\Delta_{ab}=\omega_{a}-\omega_{b}$ is the frequency deference between two atoms. 
\begin{figure}[t]
	\includegraphics[width=0.45\textwidth]{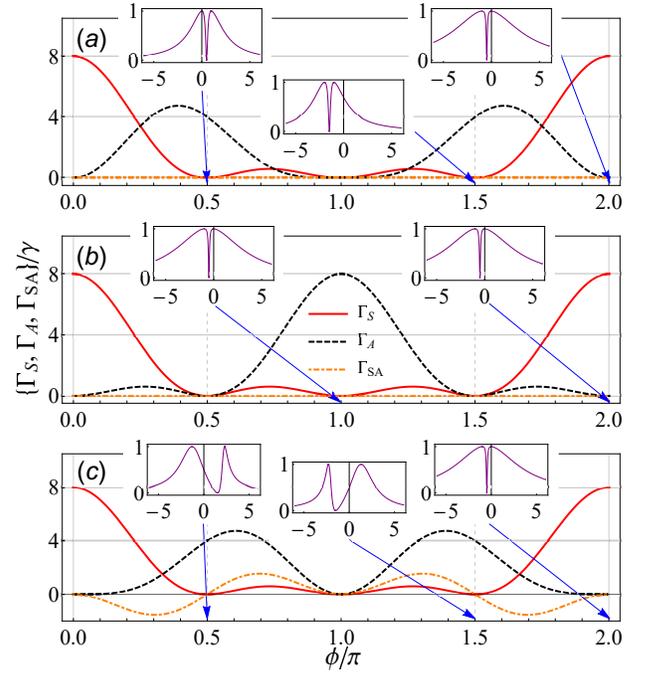}
	\caption{Individual decays $\Gamma_{\mathrm{S}}$ (red solid lines), $\Gamma_{\mathrm{A}}$ (black dashed lines), and collective decay $\Gamma_{\mathrm{SA}}$ (orange dot-dashed lines) as the functions of $\phi$ for three topologies: (a) two separate atoms, (b) two braided atoms, and (c) two nested atoms. The insets are the reflection spectra as functions of $\Delta_{a}$ (in unit of $\gamma$) at the phase delay indicated by the blue arrows, where EIT phenomena appear. The detuning between two atoms are chosen as $\Delta_{ab}=-\gamma$ in the leftmost inset in panel (c), and $\Delta_{ab}=\gamma$ in the other insets.}
	\label{E1}
\end{figure}
Clearly, when $\phi=\left(n+1/2\right)\pi$ ($n\in\mathbb{N}$), the state $|\mathrm{A}\rangle$ ($|\mathrm{S}\rangle$) become a bright (dark) state, with $\Gamma_{\mathrm{S}}=0$, $\Gamma_{\mathrm{A}}=4\gamma$, $\Gamma_{\mathrm{SA}}=0$, and
$\Delta_{\mathrm{S}}=\Delta_{\mathrm{A}}=\left(\Delta_{a}+\Delta_{b}\right)/2+(-1)^{n+1}\gamma$. The 
corresponding transmission and reflection amplitudes can be described by Eqs.~\eqref{transmissionEIT-Amode}-\eqref{reflectionEIT-Amode}. In addition, from \eqref{EITregimeSA-driveA} and Eqs.~\eqref{gSA1}, we can see that $\Delta_{ab}$ 
plays the role of control field, and the EIT regime is $0<|\Delta_{ab}|<2\gamma$. The corresponding reflection 
spectra as functions of $\Delta_{a}$ are shown in the two insets on the left (with $\phi=0.5\pi,\Delta_{ab}=\gamma$ and $\phi=1.5\pi,\Delta_{ab}=\gamma$, respectively) in Fig.~\ref{E1}(a), with transparency points located at $\Delta_{a}=(-1)^{n}\gamma-\Delta_{ab}/2$. 

On the contrary, when $\phi=2n\pi$ ($n\in\mathbb{N}^+$), the state $|\mathrm{S}\rangle$ ($|\mathrm{A}\rangle$) becomes a bright (dark) state, with $\Gamma_{\mathrm{A}}=0$, $\Gamma_{\mathrm{S}}=8\gamma$, $\Gamma_{\mathrm{SA}}=0$ and $\Delta_{\mathrm{S}}=\Delta_{\mathrm{A}}=\left(\Delta_{a}+\Delta_{b}\right)/2$. Correspondingly, the transmission and reflection amplitudes can be obtained from Eqs.~\eqref{transmissionEIT-Amode} and \eqref{reflectionEIT-Amode} by the index exchange
$\mathrm{S}\leftrightarrow\mathrm{A}$. Also, in this case, $\Delta_{ab}$ plays the role of control field, and the EIT regime is $0<|\Delta_{ab}|<4\gamma$. The corresponding reflection 
spectra are shown in the right insets (with $\phi=2\pi,\Delta_{ab}=\gamma$) in Fig.~\ref{E1}(a), with transparency point located at $\Delta_{a}=-\Delta_{ab}/2$. 

\subsubsection{Two braided giant atoms}
For two braided giant atoms with maximum symmetry, from Eqs.~\eqref{Gamma S}-\eqref{Gamma SA}, we have
\begin{subequations}
\begin{equation}
g_{\mathrm{SA}}=\frac{1}{2}\Delta_{ab},
\label{gSA2}
\end{equation}
\begin{equation}
\Gamma_{\mathrm{A}}=\gamma\left(2-3\cos\phi+2\cos 2\phi-\cos 3\phi\right),
\label{GammaA2}
\end{equation}
\begin{equation}
\Gamma_{\mathrm{S}}=\gamma\left(2+3\cos\phi+2\cos 2\phi+\cos 3\phi\right).
\label{GammaS2}
\end{equation}
\begin{equation}
\Gamma_{\mathrm{SA}}=0.
\label{GammaSA2}
\end{equation}
\end{subequations}
Obviously, when $\phi=\left(2n+1\right)\pi$ ($n\in\mathbb{N}$), the state $|\mathrm{A}\rangle$ ($|\mathrm{S}\rangle$) is the bright (dark) state, with $\Gamma_{\mathrm{S}}=0$, $\Gamma_{\mathrm{A}}=8\gamma$. And when $\phi=2n\pi$ ($n\in\mathbb{N}^+$), the state $|\mathrm{S}\rangle$ ($|\mathrm{A}\rangle$) is the bright (dark) state, with 
$\Gamma_{\mathrm{A}}=0$, $\Gamma_{\mathrm{S}}=8\gamma$. Additionally, in both cases, we have $\Delta_{\mathrm{S}}=\Delta_{\mathrm{A}}=\left(\Delta_{a}+\Delta_{b}\right)/2$ and $\Gamma_{\mathrm{SA}}=0$. Thus the two cases give rise to the same spectra.
From  Eqs.~\eqref{EITregimeSA-driveA} and \eqref{gSA2}, we find that $\Delta_{ab}$ plays the role of control field, and 
the EIT regime is $0<|\Delta_{ab}|<4\gamma$. The corresponding reflection 
spectra are shown in the insets (with $\phi=\pi,\Delta_{ab}=\gamma$ and $\phi=2\pi,\Delta_{ab}=\gamma$, respectively) in Fig.~\ref{E1}(b), with transparency 
points located at $\Delta_{a}=-\Delta_{ab}/2$. 
\subsubsection{Two nested giant atoms}
For two nested giant atoms with maximum symmetry, from Eqs.~\eqref{Gamma S}-\eqref{Gamma SA}, we have
\begin{subequations}
\begin{equation}
g_{\mathrm{SA}}=\frac{1}{2}\Delta_{ab}+\frac{1}{2}\gamma\left(\sin\phi-\sin 3\phi\right),
\label{gSA3}
\end{equation}
\begin{equation}
\Gamma_{\mathrm{A}}=\gamma\left(2-\cos\phi-2\cos 2\phi+\cos3\phi\right),
\label{GammaA3}
\end{equation}
\begin{equation}
\Gamma_{\mathrm{S}}=\gamma\left(2+3\cos\phi+2\cos 2\phi+\cos3\phi\right),
\label{GammaS3}
\end{equation}
\begin{equation}
\Gamma_{\mathrm{SA}}=\gamma\left(\cos3\phi-\cos\phi\right).
\label{GammaSA3}
\end{equation}
\end{subequations}
Clearly, when $\phi=(n+1/2)\pi$ ($n\in\mathbb{N}$), the state $|\mathrm{A}\rangle$ ($|\mathrm{S}\rangle$) couples to the waveguide and form a bright (dark) state, with $\Gamma_{\mathrm{S}}=0$, $\Gamma_{\mathrm{A}}=4\gamma$, $\Gamma_{\mathrm{SA}}=0$, $\Delta_{\mathrm{S,A}}=\left(\Delta_{a}+\Delta_{b}\right)/2\mp(-1)^{n}\gamma$ and $g_{\mathrm{SA}}=\Delta_{ab}/2+(-1)^{n}\gamma$. The corresponding 
transmission and reflection amplitudes can be described by Eqs.~\eqref{transmissionEIT-Amode}-\eqref{reflectionEIT-Amode}. In 
addition, from Eq.~\eqref{EITregimeSA-driveA}, we find that when $n$ is even, the EIT 
regime is $-4\gamma<\Delta_{ab}<0$ ($\Delta_{ab}\neq -2\gamma$). On the other hand, when $n$ is odd, 
the EIT regime is $0<\Delta_{ab}<4\gamma$ ($\Delta_{ab}\neq 2\gamma$). Note that in this case, the EIT spectrum is asymmetric because $\Delta_{\mathrm{S}}\neq\Delta_{\mathrm{A}}$. 
The corresponding reflection 
spectra are shown in the two insets on the left (with $\phi=0.5\pi, \Delta_{ab}=-\gamma$ and $\phi=1.5\pi, \Delta_{ab}=\gamma$) in Fig.~\ref{E1}(c),
with transparency points appearing at $\Delta_{a}=(-1)^{n}\gamma-\Delta_{ab}/2$.
 
When $\phi=2n\pi$ ($n\in\mathbb{N}^+$), the state $|\mathrm{S}\rangle$ ($|\mathrm{A}\rangle$) couples to the waveguide and form a bright (dark) state, with $\Gamma_{\mathrm{A}}=0$, $\Gamma_{\mathrm{S}}=8\gamma$, $\Gamma_{\mathrm{SA}}=0$, $\Delta_{\mathrm{S}}=\Delta_{\mathrm{A}}=\left(\Delta_{a}+\Delta_{b}\right)/2$, and $g_{\mathrm{SA}}=\Delta_{ab}/2$. The corresponding transmission and reflection amplitudes can be obtained from Eqs.~\eqref{transmissionEIT-Amode} and \eqref{reflectionEIT-Amode} by the index exchange
$\mathrm{S}\leftrightarrow\mathrm{A}$. The EIT 
regime is $0<|\Delta_{ab}|<4\gamma$. The corresponding reflection 
spectra are shown in the right inset (with $\phi=2\pi,\Delta_{ab}=\gamma$) in Fig.~\ref{E1}(c), with transparency located at $\Delta_{a}=-\Delta_{ab}/2$.

\subsection{\label{EITa-b}Realizing EIT using waveguide-mediated interactions between single-atom states}
\begin{figure}[t]
\includegraphics[width=0.5\textwidth]{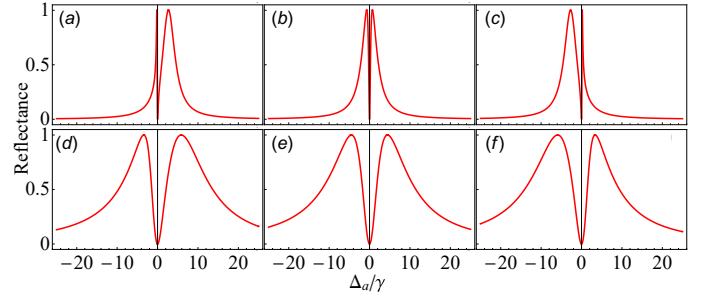}
\caption{EIT-like spectra caused by waveguide-mediated interactions between single-atom states. (a)-(c) Two braided atoms with parameters $\phi_{a1}=0$, $\phi_{a2}=\pi$, $\phi_{b1}=0.25\pi$, $\phi_{b2}=2.25\pi$, and $\gamma_{a}=\gamma_{b}=\gamma$. The detunings between two atoms are (a) $\Delta_{ab}=\Delta_{\mathrm{L},b}-2.5\gamma$, (b) $ \Delta_{ab}=\Delta_{\mathrm{L},b}$, and (c) $\Delta_{ab}=\Delta_{\mathrm{L},b}+2.5\gamma$, respectively. (d)-(f) Two nested atoms with parameters $\phi_{a1}=0$, $\phi_{a2}=\pi$, $\phi_{b1}=0.25\pi$, $\phi_{b2}=0.75\pi$, $\gamma_{a}=\gamma$ and $\gamma_{b}=10\gamma$. The detunings between two atoms are: (d) $\Delta_{ab}=\Delta_{\mathrm{L},b}-2.5\gamma$, (e) $ \Delta_{ab}=\Delta_{\mathrm{L},b}$, and (f) $\Delta_{ab}=\Delta_{\mathrm{L},b}+2.5\gamma$, respectively.}
\label{E2}
\end{figure}
In previous subsection, we discuss the EIT phenomena caused by the waveguide-mediated 
interactions between collective symmetric and anti-symmetric states. Here we investigate a different 
way to achieve EIT by engineering the waveguide-mediated interactions between single-atom states.
To this end, the following conditions should be 
satisfied: (i) one of the atoms is coupled to the waveguide, and the other is decoupled, their excitation states ($|eg\rangle$ or $|ge\rangle$) work as bright and dark states, respectively; (ii) the collective 
decay $\Gamma_{ab}$ is zero; (iii) the waveguide-mediated interaction $g_{ab}$ between the atoms is non-zero and plays a role of
control field. For example, if $\Gamma_{a}=0$, $\Gamma_{b}\neq0$, $\Gamma_{ab}=0$, and $g_{ab}\neq0$, the equation of motion Eq.~\eqref{master eq} and the Hamiltonian Eq.~\eqref{Hdrive} become
\begin{equation}
\dot{\hat{\rho}}=
-i\left[{\hat{H}_{\mathrm{drive}},\hat{\rho}}\right]+\Gamma_{b}\mathcal{D}\left[\hat{\sigma}_{b}^{-}\right]\hat{\rho},
\label{master-eq-EIT-drive-b}
\end{equation}
with
\begin{eqnarray}
\hat{H}_{\mathrm{drive}}&=&-\sum_{j}\left(\Delta_{j}-\Delta_{\mathrm{L},j}\right)\hat{\sigma}_{j}^{+}\hat{\sigma}_{j}^{-}+g_{ab}\left(\hat{\sigma}_{a}^{+}\hat{\sigma}_{b}^{-}+\hat{\sigma}_{b}^{+}\hat{\sigma}_{a}^{-}\right)
\nonumber \\
&&-\frac{i}{2}\left(\Omega_{b}\hat{\sigma}_{b}^{+}-\mathrm{H.c.}\right).
\label{Hamiltonian-EIT-drive-b}
\end{eqnarray}
Here $j=a,b$. Similarly, we can prove that above equation of motion can generate EIT or ATS type scattering spectra by
mapping it to a driven $\Lambda$-type atom. And the corresponding relations between the two systems can be summarized as follows (see Appendix~\ref{mapping-to-Lambda}): $|gg\rangle\leftrightarrow|0\rangle$, $|eg\rangle\leftrightarrow|1\rangle$, $|ge\rangle\leftrightarrow|2\rangle$, $\hat{\sigma}_{a}^{-}\leftrightarrow\hat{\sigma}_{01}$, $\hat{\sigma}_{b}^{-}\leftrightarrow\hat{\sigma}_{02}$, $g_{ab}\leftrightarrow\Omega_{\mathrm{c}}/2$, $\Omega_{b}\leftrightarrow\Omega_{\mathrm{p}}$,
$\Delta_{a}-\Delta_{\mathrm{L},a}\leftrightarrow\Delta_{\mathrm{p}}-\Delta_{\mathrm{c}}$, $\Delta_{b}-\Delta_{\mathrm{L},b}\leftrightarrow\Delta_{\mathrm{p}}$, $\Gamma_{b}\leftrightarrow\Gamma_{20}$. Note that for this case, these mappings are accurate only in the single-photon sector, as discussed in Sec.~\ref{comparison} and Appendix~\ref{ISP}.

These analysis can be
verified by simplifying Eqs.~\eqref{transmission}-\eqref{reflection} under conditions $\Gamma_{a}=0$, $\Gamma_{b}\neq0$, $\Gamma_{ab}=0$, and $g_{ab}\neq0$. The corresponding 
transmission and reflection amplitudes can be expressed as
\begin{subequations}
\begin{equation}
t=\frac{-\left(\Delta_{a}-\Delta_{\mathrm{L},a}\right)\left(\Delta_{b}-\Delta_{\mathrm{L},b}\right)+g_{ab}^{2}}{i \left(\Delta_{a}-\Delta_{\mathrm{L},a}\right)\left[i\left(\Delta_{b}-\Delta_{\mathrm{L},b}\right)-\frac{1}{2}\Gamma_{b}\right]+g_{ab}^{2}},
\label{transmissionEIT-drive-b}
\end{equation}
\begin{equation}
r=\frac{\frac{1}{2}i\Gamma_{b}\left(\Delta_{a}-\Delta_{\mathrm{L},a}\right)} {i \left(\Delta_{a}-\Delta_{\mathrm{L},a}\right)\left[i\left(\Delta_{b}-\Delta_{\mathrm{L},b}\right)-\frac{1}{2}\Gamma_{b}\right]+g_{ab}^{2}}.
\label{reflectionEIT-drive-b}
\end{equation}
\end{subequations}
By taking $\Delta_{a}-\Delta_{\mathrm{L},a}\simeq\Delta_{b}-\Delta_{\mathrm{L},b}$ and replacing them with $\mathcal{Z}$, we can inspect the complex roots of the denominator of the scattering amplitudes
\begin{equation}
\mathcal{Z}_{\pm}=-i\frac{\Gamma_{b}}{4}\pm\frac{1}{4}\sqrt{16g_{ab}^{2}-\Gamma_{b}^{2}},
\end{equation}
which are purely imaginary for
\begin{equation}
|g_{ab}|<\frac{\Gamma_{b}}{4}.
\label{EITregime-ab--drive-b}
\end{equation}
In this parameter regime, the transmission point located at $\Delta_{a}=\Delta_{\mathrm{L},a}$ is caused by 
quantum interference. This regime is so called EIT regime.

Alternatively, one can also let $\Gamma_{a}\neq0$, $\Gamma_{b}=0$, $\Gamma_{ab}=0$, and 
$g_{ab}\neq 0$, to achieve EIT-like phenomenon. All the results for this case can be obtained from 
Eqs.~\eqref {master-eq-EIT-drive-b}-\eqref{EITregime-ab--drive-b} by the index exchange
$a\leftrightarrow b$. 

As specific examples, we consider the special case with
 $\gamma_{a1}=\gamma_{a2}=\gamma_{a}$, $\gamma_{b1}=\gamma_{b2}=\gamma_{b}$. 
And without loss of generality, we let $\phi_{a1}=0$. In the follows, we will discuss the EIT-like spectra under above assumptions for different configurations.
 
Firstly, we consider the topology with two 
separate giant atoms. As discussed before, to generate EIT requires that one of the atoms is decoupled from the waveguide. Specifically, if the atom $a$ is decoupled with $\Gamma_{a}=0$, it can be seen from Eq.~\eqref{ID} that the condition $\phi_{a2}-\phi_{a1}=\left(2n+1\right)\pi$ 
($n\in\mathbb{N}$) is required. 
By using Eq.~\eqref{EI}, one can further find that under this condition, the exchange interaction $g_{ab}=0$ (i.e., vanished 
control field) always holds. Note that some detailed analysis on the relation between the individual decay and the exchange interaction can be found in Ref.~\cite{Kockum-PRL2018}. Similarly, if $\Gamma_{b}=0$, we can also obtain a 
vanished exchange interaction. Thus, in this topological configuration, the EIT-like phenomenon can not be generated based on 
waveguide-mediated interactions between single-atom states.

For two braided atoms, if we let the atom $a$ decouple from the waveguide, satisfying the condition $\phi_{a2}-\phi_{a1}=\left(2n+1\right)\pi$ ($n\in\mathbb{N}$), we can obtain from Eqs.~\eqref{LS}-\eqref{CD} that 
 $\Gamma_{a}=\Gamma_{ab}=\Delta_{\mathrm{L},a}=0$. $\Delta_{\mathrm{L},b}=\gamma_{b}\sin(\phi_{b2}-\phi_{b1})$, $\Gamma_{b}=2\gamma_{b}[1+\cos(\phi_{b2}-\phi_{b1})]$, and $g_{ab}=\sqrt{\gamma_{a}\gamma_{b}}\sin\phi_{b1}$, respectively. Thus the corresponding transmission and reflection amplitudes can be 
described by Eqs.~\eqref{transmissionEIT-drive-b} and \eqref{reflectionEIT-drive-b}. Note that the parameters should be 
appropriately chosen to satisfy $\Gamma_{b}\neq 0$ and the EIT condition Eq.~\eqref{EITregime-ab--drive-b}. We plot the reflection spectrum as a function of $\Delta_{a}$ with different $\Delta_{ab}$ in 
Figs.~\ref{E2}(a)-\ref{E2}(c). When $\Delta_{ab}=\Delta_{\mathrm{L},b}$, the spectrum is symmetric about $\Delta_{a}=0$, as shown in Fig.~\ref{E2}(b).
Alternatively, similar EIT-like spectra can be obtained when the atom $b$ is decoupled from the waveguide (not shown here).

For two nested atoms, only when the atom $a$ (the outer one) is decoupled to the waveguide with $\phi_{a2}-\phi_{a1}=\left(2n+1\right)\pi$ ($n\in\mathbb{N}$), one can obtain the EIT spectra.  And from Eqs.~\eqref{LS}-\eqref{CD}, we can obtain $\Gamma_{a}=\Gamma_{ab}=\Delta_{\mathrm{L},a}=0$, $\Delta_{\mathrm{L},b}=\gamma_{b}\sin(\phi_{b2}-\phi_{b1})$, $\Gamma_{b}=2\gamma_{b}[1+\cos(\phi_{b2}-\phi_{b1})]$, and $g_{ab}=\sqrt{\gamma_{a}\gamma_{b}}(\sin\phi_{b1}+\sin\phi_{b2})$, respectively. The corresponding transmission and reflection amplitudes can be be described by Eqs.~\eqref{transmissionEIT-drive-b} and \eqref{reflectionEIT-drive-b}.
Also, the parameters should be appropriately chosen to satisfy $\Gamma_{b}\neq 0$ and the EIT condition Eq.~\eqref{EITregime-ab--drive-b}. We plot the reflection coefficient as a function of $\Delta_{a}$ with different $\Delta_{ab}$, as shown
 in  Figs.~\ref{E2}(d)-\ref{E2}(f). When $\Delta_{ab}=\Delta_{\mathrm{L},b}$, the spectrum is symmetric about $\Delta_{a}=0$, as shown in Fig.~\ref{E2}(e). Particularly, when $\phi_{b1}=\phi_{b2}$, this system reduces to the configuration with a giant atom containing a small atom, which has been discussed in Ref.~\cite{Ask-arXiv2020}. Similar phenomenon can also be found in wQED system with two small atoms in front of a mirror~\cite{Wen-PhysRevLett.2019}, which is very close to the setup of nested giant atoms. 
 
\subsection{\label{comparison} Comparison between the two type of EIT-like effects}

In previous subsections, the expressions of EIT-type spectra are obtained by solving the one-photon scattering problem. Here, by using the method
provided in Ref. \cite{Ask-arXiv2020}, we make a 
comparison between the two type of EIT-like effects, and determine whether the transparency is a genuine EIT effect or not by checking the inelastic scattering property beyond the one-photon sector. 

For the case discussed in Sec.~\ref{S-A}, the dark state ($|\mathrm{S}\rangle$ or $|\mathrm{A}\rangle$ state, 
dependent on the parameters) is not coupled to the doubly excited state  $|ee\rangle$. Thus  $|ee\rangle$  is not occupied in the steady state [which is a dark steady state being a superposition of $|gg\rangle$ and $|\mathrm{S}\rangle$ (or $|\mathrm{A}\rangle$)] when the system 
is driven at its EIT frequency \cite{Ask-arXiv2020}. Thus the setup behaves like a proper $\Lambda$ system even 
when a multiple-photon state incident. Consequently, the fluorescence is fully quenched
at the transparency point, and the corresponding inelastic photon flux is zero. The transparency can be explained as a genuine EIT effect.  
However, for the case discussed in Sec.~\ref{EITa-b}, which is based on the exchange interaction between single-atom states, the situation is 
different. The obtained expressions of EIT-like spectra are valid only when a single-photon Fock state is incident. But if a photon state containing 
multiple-photon 
components is incident, the steady-state  occupation probability for the doubly excited state $|ee\rangle$ is nonzero because the ``dark" state 
in the single-excitation regime ($|ge\rangle$ or $|eg\rangle$, dependent on the parameters) is now coupled to $|ee\rangle$, making this setup an effective $N$-type four-level system, not a standard $\Lambda$ system.  Consequently, the fluorescence is not quenched at 
the  EIT frequency, and the corresponding inelastic photon flux is nonzero. Thus for this case the EIT effect breaks down outside the single-photon sector.

To show these in more detail, we calculate numerically the scattering coefficients and the total inelastic photon flux for these two cases based on master-equation method and provide further discussions in Appendix \ref {ISP}.

\section{\label{conclusion}CONCLUSIONS AND DISCUSSIONS}

In summary, we obtain the general analytical solutions for the single-photon scattering problem in double-giant-atom wQED systems. Our results are unified descriptions of the scattering amplitudes for three basic topologies. Using the analytical expressions of scattering spectra, we further investigate the phenomena of Fano interference and EIT without control field in these systems. 
On one hand, we discuss in detail the influences of the topological configurations and the phase delays on the Fano-like lineshapes. Typically, we show that the scattering spectrum can be used to characterize the decoherence-free interaction \cite{Kockum-PRL2018}, which is a unique phenomenon in the giant-atom structures. On the other hand, we summarize the conditions for EIT without control field in the wQED systems with two giant atoms, and verify these conditions by checking the corresponding scattering spectra. These conditions may be useful references for the future experiments on EIT-like phenomenon in the giant-atom systems. These phenomena may provide powerful tools for  controlling and manipulating photon transport in the future quantum networks.
\begin{acknowledgments}
This work was supported by the National Natural Science Foundation of China (NSFC) under Grants No. 11404269, No. 
61871333, and No. 12047576.
\end{acknowledgments}
\appendix
\section{\label{DefinitionC}Definition of the coefficients in Eq.~\eqref {Fanoshape r3}}
The moduli and the arguments of the coefficients $\chi_{\pm}$ in Eq.~\eqref{Fanoshape r3} are
\begin{subequations}
\begin{equation}
\left|\chi_{\pm}\right|=\chi=\sqrt{\left\{\frac{\gamma}{\Gamma_{+}\Gamma_{-}}\left[\lambda_{1}\left(\Gamma_{+}-\Gamma_{-}\right)+\lambda_{2}\left(\Delta_{+}-\Delta_{-}\right)\right]\right\}^{2}+\lambda_{3}^{2}},
\end{equation}
\begin{equation}
\arg{\left[\chi_{\pm}\right]}=\phi-\zeta\pm\vartheta,
\end{equation}
\end{subequations}
where
\begin{subequations}
\begin{equation}
\lambda_{1}=-\frac{1}{4\sqrt{2A}}\left|\cos\frac{\phi}{2}\right|\left(5\sin 3\phi-2\sin 4\phi+\sin 5\phi\right),
\end{equation}
\begin{equation}
\lambda_{2}=\sqrt{\frac{2}{A}}\left|\cos\frac{\phi}{2}\right|^{3}\left(2\cos\phi-\cos 2\phi-2\right)^{2},
\end{equation}
\begin{equation}
\lambda_{3}=\sqrt{\frac{2}{A}}\left|\cos\frac{\phi}{2}\right|\left(2\cos\phi-\cos 2\phi-2\right),
\end{equation}
\begin{equation}
\tan2\zeta=\frac{2\sin\phi}{1-3\cos\phi},
\end{equation}
\begin{equation}
\tan\vartheta=\frac{\lambda_{3}\Gamma_{+}\Gamma_{-}}{\left[\lambda_{1}\left(\Gamma_{+}-\Gamma_{-}\right)+\lambda_{2}\left(\Delta_{+}-\Delta_{-}\right)\right]\gamma},
\end{equation}
\begin{equation}
A=\sqrt{\left(1-3\cos\phi\right)^{2}+4\sin^{2}\phi}.
\end{equation}
\end{subequations}
\section{\label{mapping-to-Lambda} Comparison with a driven $\Lambda$-type atom}
Here we consider a three-level $\Lambda$-type atom with a ground state $|0\rangle$, a metastable state  $|1\rangle$, and an excited  state  $|2\rangle$. Only the transitions $|0\rangle  \leftrightarrow |2\rangle $ and $|1\rangle \leftrightarrow|2\rangle $ are allowed, with transition 
frequencies $\omega_{20}$ and $\omega_{21}$, respectively. To generate EIT effect, the transition $|0\rangle  \leftrightarrow |2\rangle$ is coupled by a probe field with amplitude $\Omega_{\mathrm{p}}$ and frequency $\omega_{\mathrm{p}}$, and  the transition $|1\rangle  \leftrightarrow |2\rangle$ is coupled by
a control field with amplitude $\Omega_{\mathrm{c}}$ and frequency  $\omega_{\mathrm{c}}$, respectively. The master equation describing the system dynamics can be written as \cite{Fleischhauer-Rev.Mod.Phys.2005}
\begin{equation}
\dot{\hat{\rho}}=
-i\left[{\hat{H},\hat{\rho}}\right]+\Gamma_{20}\mathcal{D}\left[\hat{\sigma}_{02}\right]\hat{\rho}+\Gamma_{21}\mathcal{D}\left[\hat{\sigma}_{12}\right]\hat{\rho},
\label{Lambda master equation}
\end{equation}
where $\hat{\sigma}_{ij}=|i\rangle \langle j|$ is the atomic transition operator. $\Gamma_{ij}$ is the decay rate from state $|i\rangle$ to state $|j\rangle $. In order to keep the physics transparent, we have ignored the pure dephasings. In a rotating frame and under RWA, the Hamiltonian of the system can be written as $(\hbar=1)$ 
\begin{equation}
\hat{H}=
-\Delta_{\mathrm{p}}\hat{\sigma}_{22}-(\Delta_{\mathrm{p}}-\Delta_{\mathrm{c}})\hat{\sigma}_{11}-\frac{1}{2}i\left(\Omega_{\mathrm{p}}\hat{\sigma}_{20}+\Omega_{\mathrm{c}}\hat{\sigma}_{21}-\mathrm{H.c.}\right),
\label{Lambda H}
\end{equation}
where $\Delta_{\mathrm{p}}=\omega_{\mathrm{p}}-\omega_{20}$ and $\Delta_{\mathrm{c}}=\omega_{\mathrm{c}}-\omega_{21}$ are the detunings of the probe and the control fields, respectively. If the driving fields are applied through a wQED structure, the transmission and reflection amplitudes under the weak-probe limit $\Omega_{\mathrm{p}}\ll\Omega_{\mathrm{c}},\Gamma_{20}$ can be expressed as 
\begin{subequations}
\begin{equation}
t=\frac{i\left(\Delta_{\mathrm{p}}-\Delta_{\mathrm{c}}\right)\left(i\Delta_{\mathrm{p}}-\frac{1}{2}\Gamma_{21}\right)+\frac{1}{4}\Omega_{\mathrm{c}}^{2}} {i \left(\Delta_{\mathrm{p}}-\Delta_{\mathrm{c}}\right)\left[i\Delta_{\mathrm{p}}-\frac{1}{2}\left(\Gamma_{20}+\Gamma_{21}\right)\right]+\frac{1}{4}\Omega_{\mathrm{c}}^{2}},
\label{transmissionEIT-Lambda}
\end{equation}
\begin{equation}
r=\frac{\frac{1}{2}i\Gamma_{20}\left(\Delta_{\mathrm{p}}-\Delta_{\mathrm{c}}\right)} {i \left(\Delta_{\mathrm{p}}-\Delta_{\mathrm{c}}\right)\left[i\Delta_{\mathrm{p}}-\frac{1}{2}\left(\Gamma_{20}+\Gamma_{21}\right)\right]+\frac{1}{4}\Omega_{\mathrm{c}}^{2}}.\label{reflectionEIT-Lambda}
\end{equation}
\end{subequations}

By comparing above results with the master equation \eqref{master-eq-EIT-driveA}, the Hamiltonian \eqref{Hamiltonian-EIT-driveA}, and the 
scattering amplitudes \eqref{transmissionEIT-Amode} and  \eqref{reflectionEIT-Amode} [where  
the state $|\mathrm{S}\rangle$ ($|\mathrm{A}\rangle$) plays the role of the dark (bright) state], and assuming $\Gamma_{21}=0$, we can make the identifications $|gg\rangle\leftrightarrow|0\rangle$, $|\mathrm{S}\rangle\leftrightarrow|1\rangle$, $|\mathrm{A}\rangle\leftrightarrow|2\rangle$, $\hat{\sigma}_{\mathrm{S}}^{-}\leftrightarrow\hat{\sigma}_{01}$, $\hat{\sigma}_{\mathrm{A}}^{-}\leftrightarrow\hat{\sigma}_{02}$, $g_{\mathrm{SA}}\leftrightarrow\Omega_{\mathrm{c}}/2$, $\Omega_{\mathrm{A}}\leftrightarrow\Omega_{\mathrm{p}}$, $\Delta_{\mathrm{S}}\leftrightarrow\Delta_{\mathrm{p}}-\Delta_{\mathrm{c}}$, $\Delta_{\mathrm{A}}\leftrightarrow\Delta_{\mathrm{p}}$, $\Gamma_{\mathrm{A}}\leftrightarrow\Gamma_{20}$. Straightforwardly, for the case that the state $|\mathrm{A}\rangle$ ($|\mathrm{S}\rangle$) plays the role of the dark (bright) state, the mappings can be obtained from above results by exchanging A and S.

Similarly, for the case described by the  master equation \eqref{master-eq-EIT-drive-b} and the Hamiltonian \eqref{Hamiltonian-EIT-drive-b} [where  
the state $|eg\rangle$ ($|ge\rangle$) plays the role of the dark (bright) state], we can make the identifications: $|gg\rangle\leftrightarrow|0\rangle$, $|eg\rangle\leftrightarrow|1\rangle$, $|ge\rangle\leftrightarrow|2\rangle$, $\hat{\sigma}_{a}^{-}\leftrightarrow\hat{\sigma}_{01}$, $\hat{\sigma}_{b}^{-}\leftrightarrow\hat{\sigma}_{02}$, $g_{ab}\leftrightarrow\Omega_{\mathrm{c}}/2$, $\Omega_{b}\leftrightarrow\Omega_{\mathrm{p}}$,
$\Delta_{a}-\Delta_{\mathrm{L},a}\leftrightarrow\Delta_{\mathrm{p}}-\Delta_{\mathrm{c}}$, $\Delta_{b}-\Delta_{\mathrm{L},b}\leftrightarrow\Delta_{\mathrm{p}}$, $\Gamma_{b}\leftrightarrow\Gamma_{20}$.
For the case that the atom $b$ is decoupled, the mappings can be obtained by exchanging $a$ and $b$.

Note that above mappings are accurate only in the single-excitation subspace, where the state $|ee\rangle$ is excluded. 
For situations beyond the one-photon sector, we should check if there exists a fluorescence quench to determine whether the transparency is a genuine EIT effect or not. We discuss this issue in Sec.~\ref{comparison} and Appendix~\ref {ISP}.
\section{\label{ISP} Inelastic scattering properties under the EIT conditions}
\begin{figure}[t]
\includegraphics[width=0.45\textwidth]{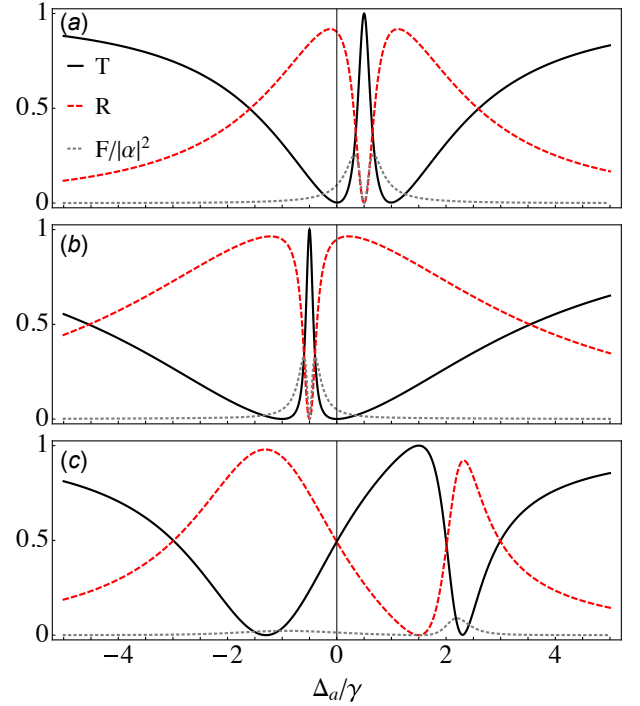}
\caption {Transmission coefficient (solid lines), reflection coefficient (dashed lines), and inelastic photon flux (dotted lines) as a function of probe detuning $\Delta_{a}$ for three different configurations. All systems are in the parameter regime where they fulfill the EIT criteria given in Sec.~\ref{S-A}. (a) Two separate giant atoms, the system parameters are the same as those used in the left inset in Fig.~\ref{E1}(a), the coherent drive amplitude is $|\alpha|^2=0.04\gamma$;  (b) Two braided giant atoms,  the system parameter are the same as those used in the left inset in Fig.~\ref{E1}(b),  the coherent drive amplitude is $|\alpha|^2=0.04\gamma$;  (c) Two nested giant atoms,  the system parameters are the same as those used in the left inset in Fig.~\ref{E1}(c),  the coherent drive amplitude is $|\alpha|^2=0.01\gamma$.}
\label{A1}
\end{figure}
\begin{figure}[t]
\includegraphics[width=0.45\textwidth]{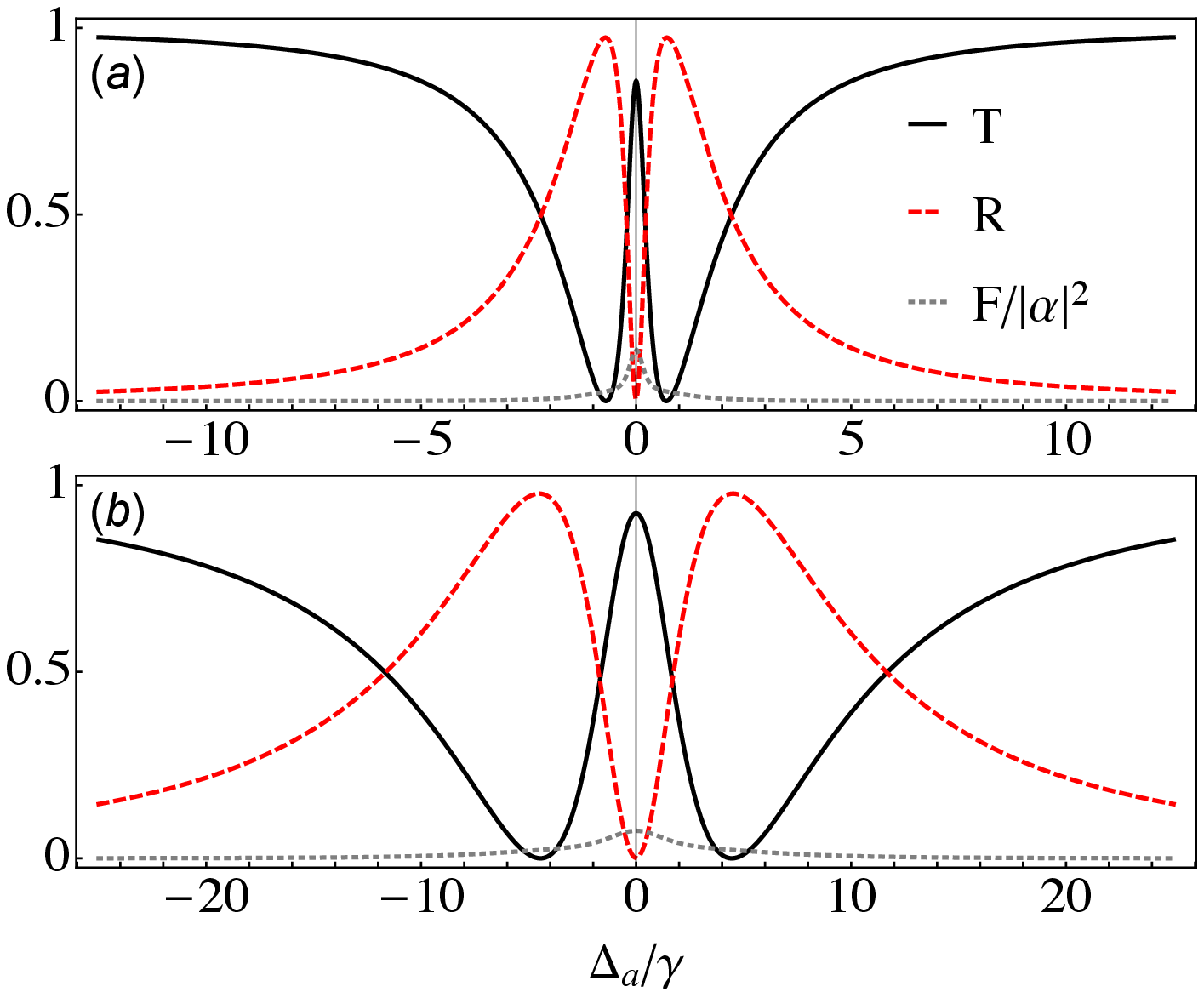}
\caption{Transmission coefficient (solid lines), reflection coefficient (dashed lines), and inelastic photon flux (dotted lines) as a function of probe detuning $\Delta_{a}$ for the braised and the nested configurations. All systems are in the parameter regime where they fulfill the EIT criteria given in Sec.~\ref{EITa-b}. (a) Two braided giant atoms,  the system parameter are the same as those used in Fig.~\ref{E2}(b),  the coherent drive amplitude is $|\alpha|^2=0.01\gamma$;  (b) Two nested giant atoms,  the system parameter are the same as those used in Fig.~\ref{E2}(e),  the coherent drive amplitude is $|\alpha|^2=0.04\gamma$.}
\label{A2}
\end{figure}
In this appendix,  we calculate numerically the scattering coefficients and the total inelastic photon flux for all the cases in Sec.~\ref{EITwithoutCF}. In our simulation, we use a weak coherent field as a probe. 
By checking  if there exists a fluorescence quench, we can verify that for the case discussed in Sec.~\ref{S-A}, the transparency can be explained as a genuine EIT effect, but for the case in Sec.~\ref{EITa-b}, the EIT effect breaks down outside the single-photon sector. 

Using input-output theory, the output operators describing the transmission and reflection bosonic fields can be written as  \cite{Kockum-PRL2018, Ask-arXiv2020}
\begin{subequations}
\begin{equation}
\hat{b}^{(\mathrm{t})}_{\mathrm{out}}=\alpha e^{i\phi_{N^{'},a1}}+\sum_{jn}e^{i\phi_{N^{'},jn}}\sqrt{\frac{\gamma_{jn}}{2}}\hat{\sigma}^{-}_{j},
\end{equation}
\begin{equation}
\hat{b}^{(\mathrm{r})}_{\mathrm{out}}=\sum_{jn}e^{i\phi_{jn,a1}}\sqrt{\frac{\gamma_{jn}}{2}}\hat{\sigma}^{-}_{j},
\end{equation}
\end{subequations}
where $j=a,b$ and $n=1,2$. $N'$ is used to label the rightmost coupling point. 
The transmission and reflection amplitudes can be further defined as 
\begin{subequations}
\begin{equation}
t=\frac{\langle\hat{b}^{(\mathrm{t})}_{\mathrm{out}}\rangle}{\alpha}=e^{i\phi_{N^{'},a1}}+\frac{1}{\alpha}\sum_{jn}e^{i\phi_{N^{'},jn}}\sqrt{\frac{\gamma_{jn}}{2}}\langle\hat{\sigma}^{-}_{j}\rangle,
\end{equation}
\begin{equation}
r=\frac{\langle\hat{b}^{(\mathrm{r})}_{\mathrm{out}}\rangle}{\alpha}=\frac{1}{\alpha}\sum_{jn}e^{i\phi_{jn,a1}}\sqrt{\frac{\gamma_{jn}}{2}}\langle\hat{\sigma}^{-}_{j}\rangle,
\end{equation}
\end{subequations}
where $\langle\hat{\sigma}^{-}_{j}\rangle=\mathrm{Tr}[\hat{\rho}\hat{\sigma}^{-}_{j}]$ is the steady-state expectation value of lower operator $\hat{\sigma}^{-}_{j}$, which can be obtained by numerically solving the master equation \eqref{master eq}. The corresponding transmission
and reflection coefficients are $T=|t|^2$ and $R=|r|^2$.

To study the inelastic scattering properties, we define the total inelastic photon flux \cite{Fang-PRA2015}
\begin{equation}
F(\omega)=
\sum_{\mathrm{i}=\mathrm{t},\mathrm{r}}\int{S_{\omega}^{(\mathrm{i})}(\nu)\mathrm{d}\nu},
\end{equation} 
where 
\begin{equation}
S_{\omega}^{\mathrm{(i)}}(\nu)=\int{e^{-i\nu t}}\langle\hat{b}_{\mathrm{out}}^{(\mathrm{i})\dag}(t)\hat{b}_{\mathrm{out}}^{(\mathrm{i})}(0)\rangle\mathrm{d}t
\end{equation} 
($\mathrm{i}=\mathrm{t},\mathrm{r}$) is the inelastic power spectrum when the system is driven by a coherent field with frequency $\omega$. The steady-state correlation function $\langle\hat{b}_{\mathrm{out}}^{(\mathrm{i})\dag}(t)\hat{b}_{\mathrm{out}}^{(\mathrm{i})}(0)\rangle$ can be calculated 
using the solution to the master equation \eqref{master eq}. 

We plot the transmittance $T$, the reflectance $R$, and the flux $F$ as a function of probe detuning $\Delta_{a}$ in the EIT regime in 
Figs.~\eqref{A1} and \eqref{A2}. The results show that these quantities satisfy relation $F/|\alpha|^2=1-T-R$, showing that photon-number 
conservation is preserved. For EIT based on atomic collective states (the case discussed in Sec.~\ref{S-A}), we can find that for each 
configuration, the inelastic photon flux $F$ is zero (i.e., the fluorescence is quenched) and the total transparency is preserved at the EIT 
frequency  [Figs.~\eqref{A1}(a)-\eqref{A1}(c)].  Thus the transparency in this case can be explained as a genuine EIT effect. On the contrary, for 
EIT-like phenomenon based on single-atom states (the case discussed in Sec.~\ref{EITa-b}), the flux $F$ is nonzero at the transparency 
frequency (i.e. the fluorescence is not quenched), which means that the inelastic scattering occurs [Figs.~\eqref{A2}(a)-\eqref{A2}(b)]. Thus for 
this case, the EIT effect breaks down when
the system is driven by a coherent field containing multi-photon components.
\bibliography{SL-Nov-30-2021}

\providecommand{\noopsort}[1]{}\providecommand{\singleletter}[1]{#1}%
\begin{thebibliography}{75}%
\makeatletter
\providecommand \@ifxundefined [1]{%
 \@ifx{#1\undefined}
}%
\providecommand \@ifnum [1]{%
 \ifnum #1\expandafter \@firstoftwo
 \else \expandafter \@secondoftwo
 \fi
}%
\providecommand \@ifx [1]{%
 \ifx #1\expandafter \@firstoftwo
 \else \expandafter \@secondoftwo
 \fi
}%
\providecommand \natexlab [1]{#1}%
\providecommand \enquote  [1]{``#1''}%
\providecommand \bibnamefont  [1]{#1}%
\providecommand \bibfnamefont [1]{#1}%
\providecommand \citenamefont [1]{#1}%
\providecommand \href@noop [0]{\@secondoftwo}%
\providecommand \href [0]{\begingroup \@sanitize@url \@href}%
\providecommand \@href[1]{\@@startlink{#1}\@@href}%
\providecommand \@@href[1]{\endgroup#1\@@endlink}%
\providecommand \@sanitize@url [0]{\catcode `\\12\catcode `\$12\catcode
  `\&12\catcode `\#12\catcode `\^12\catcode `\_12\catcode `\%12\relax}%
\providecommand \@@startlink[1]{}%
\providecommand \@@endlink[0]{}%
\providecommand \url  [0]{\begingroup\@sanitize@url \@url }%
\providecommand \@url [1]{\endgroup\@href {#1}{\urlprefix }}%
\providecommand \urlprefix  [0]{URL }%
\providecommand \Eprint [0]{\href }%
\providecommand \doibase [0]{http://dx.doi.org/}%
\providecommand \selectlanguage [0]{\@gobble}%
\providecommand \bibinfo  [0]{\@secondoftwo}%
\providecommand \bibfield  [0]{\@secondoftwo}%
\providecommand \translation [1]{[#1]}%
\providecommand \BibitemOpen [0]{}%
\providecommand \bibitemStop [0]{}%
\providecommand \bibitemNoStop [0]{.\EOS\space}%
\providecommand \EOS [0]{\spacefactor3000\relax}%
\providecommand \BibitemShut  [1]{\csname bibitem#1\endcsname}%
\let\auto@bib@innerbib\@empty
\bibitem [{\citenamefont {Roy}\ \emph {et~al.}(2017)\citenamefont {Roy},
  \citenamefont {Wilson},\ and\ \citenamefont {Firstenberg}}]{Roy-PRM2017}%
  \BibitemOpen
  \bibfield  {author} {\bibinfo {author} {\bibfnamefont {D.}~\bibnamefont
  {Roy}}, \bibinfo {author} {\bibfnamefont {C.~M.}\ \bibnamefont {Wilson}}, \
  and\ \bibinfo {author} {\bibfnamefont {O.}~\bibnamefont {Firstenberg}},\
  }\href {\doibase 10.1103/RevModPhys.89.021001} {\bibfield  {journal}
  {\bibinfo  {journal} {Rev. Mod. Phys.}\ }\textbf {\bibinfo {volume} {89}},\
  \bibinfo {pages} {021001} (\bibinfo {year} {2017})}\BibitemShut {NoStop}%
\bibitem [{\citenamefont {Gu}\ \emph {et~al.}(2017)\citenamefont {Gu},
  \citenamefont {Kockum}, \citenamefont {Miranowicz}, \citenamefont {xi~Liu},\
  and\ \citenamefont {Nori}}]{Gu-PhysReports2017}%
  \BibitemOpen
  \bibfield  {author} {\bibinfo {author} {\bibfnamefont {X.}~\bibnamefont
  {Gu}}, \bibinfo {author} {\bibfnamefont {A.~F.}\ \bibnamefont {Kockum}},
  \bibinfo {author} {\bibfnamefont {A.}~\bibnamefont {Miranowicz}}, \bibinfo
  {author} {\bibfnamefont {Y.}~\bibnamefont {xi~Liu}}, \ and\ \bibinfo {author}
  {\bibfnamefont {F.}~\bibnamefont {Nori}},\ }\href {\doibase
  https://doi.org/10.1016/j.physrep.2017.10.002} {\bibfield  {journal}
  {\bibinfo  {journal} {Physics Reports}\ }\textbf {\bibinfo {volume}
  {718-719}},\ \bibinfo {pages} {1} (\bibinfo {year} {2017})}\BibitemShut
  {NoStop}%
\bibitem [{\citenamefont {Astafiev}\ \emph {et~al.}(2010)\citenamefont
  {Astafiev}, \citenamefont {Zagoskin}, \citenamefont {Abdumalikov},
  \citenamefont {Pashkin}, \citenamefont {Yamamoto}, \citenamefont {Inomata},
  \citenamefont {Nakamura},\ and\ \citenamefont {Tsai}}]{Astafiev-Sci2010}%
  \BibitemOpen
  \bibfield  {author} {\bibinfo {author} {\bibfnamefont {O.}~\bibnamefont
  {Astafiev}}, \bibinfo {author} {\bibfnamefont {A.~M.}\ \bibnamefont
  {Zagoskin}}, \bibinfo {author} {\bibfnamefont {A.~A.}\ \bibnamefont
  {Abdumalikov}}, \bibinfo {author} {\bibfnamefont {Y.~A.}\ \bibnamefont
  {Pashkin}}, \bibinfo {author} {\bibfnamefont {T.}~\bibnamefont {Yamamoto}},
  \bibinfo {author} {\bibfnamefont {K.}~\bibnamefont {Inomata}}, \bibinfo
  {author} {\bibfnamefont {Y.}~\bibnamefont {Nakamura}}, \ and\ \bibinfo
  {author} {\bibfnamefont {J.~S.}\ \bibnamefont {Tsai}},\ }\href {\doibase
  10.1126/science.1181918} {\bibfield  {journal} {\bibinfo  {journal}
  {Science}\ }\textbf {\bibinfo {volume} {327}},\ \bibinfo {pages} {840}
  (\bibinfo {year} {2010})}\BibitemShut {NoStop}%
\bibitem [{\citenamefont {Hoi}\ \emph {et~al.}(2011)\citenamefont {Hoi},
  \citenamefont {Wilson}, \citenamefont {Johansson}, \citenamefont {Palomaki},
  \citenamefont {Peropadre},\ and\ \citenamefont {Delsing}}]{Hoi-PRL2011}%
  \BibitemOpen
  \bibfield  {author} {\bibinfo {author} {\bibfnamefont {I.-C.}\ \bibnamefont
  {Hoi}}, \bibinfo {author} {\bibfnamefont {C.~M.}\ \bibnamefont {Wilson}},
  \bibinfo {author} {\bibfnamefont {G.}~\bibnamefont {Johansson}}, \bibinfo
  {author} {\bibfnamefont {T.}~\bibnamefont {Palomaki}}, \bibinfo {author}
  {\bibfnamefont {B.}~\bibnamefont {Peropadre}}, \ and\ \bibinfo {author}
  {\bibfnamefont {P.}~\bibnamefont {Delsing}},\ }\href {\doibase
  10.1103/PhysRevLett.107.073601} {\bibfield  {journal} {\bibinfo  {journal}
  {Phys. Rev. Lett.}\ }\textbf {\bibinfo {volume} {107}},\ \bibinfo {pages}
  {073601} (\bibinfo {year} {2011})}\BibitemShut {NoStop}%
\bibitem [{\citenamefont {Shen}\ and\ \citenamefont
  {Fan}(2005{\natexlab{a}})}]{Shen-Opt.Lett.2005}%
  \BibitemOpen
  \bibfield  {author} {\bibinfo {author} {\bibfnamefont {J.-T.}\ \bibnamefont
  {Shen}}\ and\ \bibinfo {author} {\bibfnamefont {S.}~\bibnamefont {Fan}},\
  }\href {\doibase 10.1364/OL.30.002001} {\bibfield  {journal} {\bibinfo
  {journal} {Opt. Lett.}\ }\textbf {\bibinfo {volume} {30}},\ \bibinfo {pages}
  {2001} (\bibinfo {year} {2005}{\natexlab{a}})}\BibitemShut {NoStop}%
\bibitem [{\citenamefont {Shen}\ and\ \citenamefont
  {Fan}(2005{\natexlab{b}})}]{Shen-PRL2005}%
  \BibitemOpen
  \bibfield  {author} {\bibinfo {author} {\bibfnamefont {J.-T.}\ \bibnamefont
  {Shen}}\ and\ \bibinfo {author} {\bibfnamefont {S.}~\bibnamefont {Fan}},\
  }\href {\doibase 10.1103/PhysRevLett.95.213001} {\bibfield  {journal}
  {\bibinfo  {journal} {Phys. Rev. Lett.}\ }\textbf {\bibinfo {volume} {95}},\
  \bibinfo {pages} {213001} (\bibinfo {year} {2005}{\natexlab{b}})}\BibitemShut
  {NoStop}%
\bibitem [{\citenamefont {Chang}\ \emph {et~al.}(2006)\citenamefont {Chang},
  \citenamefont {S\o{}rensen}, \citenamefont {Hemmer},\ and\ \citenamefont
  {Lukin}}]{Chang-PRL2006}%
  \BibitemOpen
  \bibfield  {author} {\bibinfo {author} {\bibfnamefont {D.~E.}\ \bibnamefont
  {Chang}}, \bibinfo {author} {\bibfnamefont {A.~S.}\ \bibnamefont
  {S\o{}rensen}}, \bibinfo {author} {\bibfnamefont {P.~R.}\ \bibnamefont
  {Hemmer}}, \ and\ \bibinfo {author} {\bibfnamefont {M.~D.}\ \bibnamefont
  {Lukin}},\ }\href {\doibase 10.1103/PhysRevLett.97.053002} {\bibfield
  {journal} {\bibinfo  {journal} {Phys. Rev. Lett.}\ }\textbf {\bibinfo
  {volume} {97}},\ \bibinfo {pages} {053002} (\bibinfo {year}
  {2006})}\BibitemShut {NoStop}%
\bibitem [{\citenamefont {Shen}\ and\ \citenamefont
  {Fan}(2007)}]{Shen-PRL2007}%
  \BibitemOpen
  \bibfield  {author} {\bibinfo {author} {\bibfnamefont {J.-T.}\ \bibnamefont
  {Shen}}\ and\ \bibinfo {author} {\bibfnamefont {S.}~\bibnamefont {Fan}},\
  }\href {\doibase 10.1103/PhysRevLett.98.153003} {\bibfield  {journal}
  {\bibinfo  {journal} {Phys. Rev. Lett.}\ }\textbf {\bibinfo {volume} {98}},\
  \bibinfo {pages} {153003} (\bibinfo {year} {2007})}\BibitemShut {NoStop}%
\bibitem [{\citenamefont {Chang}\ \emph {et~al.}(2007)\citenamefont {Chang},
  \citenamefont {S\o{}rensen}, \citenamefont {Demler},\ and\ \citenamefont
  {Lukin}}]{Chang-Nature2007}%
  \BibitemOpen
  \bibfield  {author} {\bibinfo {author} {\bibfnamefont {D.~E.}\ \bibnamefont
  {Chang}}, \bibinfo {author} {\bibfnamefont {A.~S.}\ \bibnamefont
  {S\o{}rensen}}, \bibinfo {author} {\bibfnamefont {E.~A.}\ \bibnamefont
  {Demler}}, \ and\ \bibinfo {author} {\bibfnamefont {M.~D.}\ \bibnamefont
  {Lukin}},\ }\href {\doibase 10.1038/nphys708} {\bibfield  {journal} {\bibinfo
   {journal} {Nat. Phys.}\ }\textbf {\bibinfo {volume} {3}},\ \bibinfo {pages}
  {807} (\bibinfo {year} {2007})}\BibitemShut {NoStop}%
\bibitem [{\citenamefont {Zhou}\ \emph {et~al.}(2008)\citenamefont {Zhou},
  \citenamefont {Gong}, \citenamefont {Liu}, \citenamefont {Sun},\ and\
  \citenamefont {Nori}}]{Zhou-PRL2008}%
  \BibitemOpen
  \bibfield  {author} {\bibinfo {author} {\bibfnamefont {L.}~\bibnamefont
  {Zhou}}, \bibinfo {author} {\bibfnamefont {Z.~R.}\ \bibnamefont {Gong}},
  \bibinfo {author} {\bibfnamefont {Y.-x.}\ \bibnamefont {Liu}}, \bibinfo
  {author} {\bibfnamefont {C.~P.}\ \bibnamefont {Sun}}, \ and\ \bibinfo
  {author} {\bibfnamefont {F.}~\bibnamefont {Nori}},\ }\href {\doibase
  10.1103/PhysRevLett.101.100501} {\bibfield  {journal} {\bibinfo  {journal}
  {Phys. Rev. Lett.}\ }\textbf {\bibinfo {volume} {101}},\ \bibinfo {pages}
  {100501} (\bibinfo {year} {2008})}\BibitemShut {NoStop}%
\bibitem [{\citenamefont {Shi}\ and\ \citenamefont {Sun}(2009)}]{Shi-PRB2009}%
  \BibitemOpen
  \bibfield  {author} {\bibinfo {author} {\bibfnamefont {T.}~\bibnamefont
  {Shi}}\ and\ \bibinfo {author} {\bibfnamefont {C.~P.}\ \bibnamefont {Sun}},\
  }\href {\doibase 10.1103/PhysRevB.79.205111} {\bibfield  {journal} {\bibinfo
  {journal} {Phys. Rev. B}\ }\textbf {\bibinfo {volume} {79}},\ \bibinfo
  {pages} {205111} (\bibinfo {year} {2009})}\BibitemShut {NoStop}%
\bibitem [{\citenamefont {Shen}\ and\ \citenamefont
  {Fan}(2009)}]{Shen-PRA2009}%
  \BibitemOpen
  \bibfield  {author} {\bibinfo {author} {\bibfnamefont {J.-T.}\ \bibnamefont
  {Shen}}\ and\ \bibinfo {author} {\bibfnamefont {S.}~\bibnamefont {Fan}},\
  }\href {\doibase 10.1103/PhysRevA.79.023837} {\bibfield  {journal} {\bibinfo
  {journal} {Phys. Rev. A}\ }\textbf {\bibinfo {volume} {79}},\ \bibinfo
  {pages} {023837} (\bibinfo {year} {2009})}\BibitemShut {NoStop}%
\bibitem [{\citenamefont {Longo}\ \emph {et~al.}(2010)\citenamefont {Longo},
  \citenamefont {Schmitteckert},\ and\ \citenamefont {Busch}}]{Longo-PRL2010}%
  \BibitemOpen
  \bibfield  {author} {\bibinfo {author} {\bibfnamefont {P.}~\bibnamefont
  {Longo}}, \bibinfo {author} {\bibfnamefont {P.}~\bibnamefont
  {Schmitteckert}}, \ and\ \bibinfo {author} {\bibfnamefont {K.}~\bibnamefont
  {Busch}},\ }\href {\doibase 10.1103/PhysRevLett.104.023602} {\bibfield
  {journal} {\bibinfo  {journal} {Phys. Rev. Lett.}\ }\textbf {\bibinfo
  {volume} {104}},\ \bibinfo {pages} {023602} (\bibinfo {year}
  {2010})}\BibitemShut {NoStop}%
\bibitem [{\citenamefont {Zheng}\ \emph {et~al.}(2010)\citenamefont {Zheng},
  \citenamefont {Gauthier},\ and\ \citenamefont {Baranger}}]{Zheng-PRA2010}%
  \BibitemOpen
  \bibfield  {author} {\bibinfo {author} {\bibfnamefont {H.}~\bibnamefont
  {Zheng}}, \bibinfo {author} {\bibfnamefont {D.~J.}\ \bibnamefont {Gauthier}},
  \ and\ \bibinfo {author} {\bibfnamefont {H.~U.}\ \bibnamefont {Baranger}},\
  }\href {\doibase 10.1103/PhysRevA.82.063816} {\bibfield  {journal} {\bibinfo
  {journal} {Phys. Rev. A}\ }\textbf {\bibinfo {volume} {82}},\ \bibinfo
  {pages} {063816} (\bibinfo {year} {2010})}\BibitemShut {NoStop}%
\bibitem [{\citenamefont {Fan}\ \emph {et~al.}(2010)\citenamefont {Fan},
  \citenamefont {Kocaba\ifmmode~\mbox{\c{s}}\else \c{s}\fi{}},\ and\
  \citenamefont {Shen}}]{Fan-PRA2010}%
  \BibitemOpen
  \bibfield  {author} {\bibinfo {author} {\bibfnamefont {S.}~\bibnamefont
  {Fan}}, \bibinfo {author} {\bibfnamefont
  {{\ifmmode\mbox{\c{S}}\else\c{S}\fi{}}.~E.}\ \bibnamefont
  {Kocaba\ifmmode~\mbox{\c{s}}\else \c{s}\fi{}}}, \ and\ \bibinfo {author}
  {\bibfnamefont {J.-T.}\ \bibnamefont {Shen}},\ }\href {\doibase
  10.1103/PhysRevA.82.063821} {\bibfield  {journal} {\bibinfo  {journal} {Phys.
  Rev. A}\ }\textbf {\bibinfo {volume} {82}},\ \bibinfo {pages} {063821}
  (\bibinfo {year} {2010})}\BibitemShut {NoStop}%
\bibitem [{\citenamefont {Witthaut}\ and\ \citenamefont
  {S{\o}rensen}(2010)}]{Witthaut-NJP2010}%
  \BibitemOpen
  \bibfield  {author} {\bibinfo {author} {\bibfnamefont {D.}~\bibnamefont
  {Witthaut}}\ and\ \bibinfo {author} {\bibfnamefont {A.~S.}\ \bibnamefont
  {S{\o}rensen}},\ }\href {\doibase 10.1088/1367-2630/12/4/043052} {\bibfield
  {journal} {\bibinfo  {journal} {New J. Phys.}\ }\textbf {\bibinfo {volume}
  {12}},\ \bibinfo {pages} {043052} (\bibinfo {year} {2010})}\BibitemShut
  {NoStop}%
\bibitem [{\citenamefont {Roy}(2011)}]{Roy-PRL2011}%
  \BibitemOpen
  \bibfield  {author} {\bibinfo {author} {\bibfnamefont {D.}~\bibnamefont
  {Roy}},\ }\href {\doibase 10.1103/PhysRevLett.106.053601} {\bibfield
  {journal} {\bibinfo  {journal} {Phys. Rev. Lett.}\ }\textbf {\bibinfo
  {volume} {106}},\ \bibinfo {pages} {053601} (\bibinfo {year}
  {2011})}\BibitemShut {NoStop}%
\bibitem [{\citenamefont {Zheng}\ \emph {et~al.}(2011)\citenamefont {Zheng},
  \citenamefont {Gauthier},\ and\ \citenamefont {Baranger}}]{Zheng-PRL2011}%
  \BibitemOpen
  \bibfield  {author} {\bibinfo {author} {\bibfnamefont {H.}~\bibnamefont
  {Zheng}}, \bibinfo {author} {\bibfnamefont {D.~J.}\ \bibnamefont {Gauthier}},
  \ and\ \bibinfo {author} {\bibfnamefont {H.~U.}\ \bibnamefont {Baranger}},\
  }\href {\doibase 10.1103/PhysRevLett.107.223601} {\bibfield  {journal}
  {\bibinfo  {journal} {Phys. Rev. Lett.}\ }\textbf {\bibinfo {volume} {107}},\
  \bibinfo {pages} {223601} (\bibinfo {year} {2011})}\BibitemShut {NoStop}%
\bibitem [{\citenamefont {Hoi}\ \emph {et~al.}(2013)\citenamefont {Hoi},
  \citenamefont {Kockum}, \citenamefont {Palomaki}, \citenamefont {Stace},
  \citenamefont {Fan}, \citenamefont {Tornberg}, \citenamefont {Sathyamoorthy},
  \citenamefont {Johansson}, \citenamefont {Delsing},\ and\ \citenamefont
  {Wilson}}]{Hoi-PRL2013}%
  \BibitemOpen
  \bibfield  {author} {\bibinfo {author} {\bibfnamefont {I.-C.}\ \bibnamefont
  {Hoi}}, \bibinfo {author} {\bibfnamefont {A.~F.}\ \bibnamefont {Kockum}},
  \bibinfo {author} {\bibfnamefont {T.}~\bibnamefont {Palomaki}}, \bibinfo
  {author} {\bibfnamefont {T.~M.}\ \bibnamefont {Stace}}, \bibinfo {author}
  {\bibfnamefont {B.}~\bibnamefont {Fan}}, \bibinfo {author} {\bibfnamefont
  {L.}~\bibnamefont {Tornberg}}, \bibinfo {author} {\bibfnamefont {S.~R.}\
  \bibnamefont {Sathyamoorthy}}, \bibinfo {author} {\bibfnamefont
  {G.}~\bibnamefont {Johansson}}, \bibinfo {author} {\bibfnamefont
  {P.}~\bibnamefont {Delsing}}, \ and\ \bibinfo {author} {\bibfnamefont
  {C.~M.}\ \bibnamefont {Wilson}},\ }\href {\doibase
  10.1103/PhysRevLett.111.053601} {\bibfield  {journal} {\bibinfo  {journal}
  {Phys. Rev. Lett.}\ }\textbf {\bibinfo {volume} {111}},\ \bibinfo {pages}
  {053601} (\bibinfo {year} {2013})}\BibitemShut {NoStop}%
\bibitem [{\citenamefont {Jia}\ and\ \citenamefont {Wang}(2013)}]{Jia-PRA2013}%
  \BibitemOpen
  \bibfield  {author} {\bibinfo {author} {\bibfnamefont {W.~Z.}\ \bibnamefont
  {Jia}}\ and\ \bibinfo {author} {\bibfnamefont {Z.~D.}\ \bibnamefont {Wang}},\
  }\href {\doibase 10.1103/PhysRevA.88.063821} {\bibfield  {journal} {\bibinfo
  {journal} {Phys. Rev. A}\ }\textbf {\bibinfo {volume} {88}},\ \bibinfo
  {pages} {063821} (\bibinfo {year} {2013})}\BibitemShut {NoStop}%
\bibitem [{\citenamefont {Laakso}\ and\ \citenamefont
  {Pletyukhov}(2014)}]{Laakso-PRL2014}%
  \BibitemOpen
  \bibfield  {author} {\bibinfo {author} {\bibfnamefont {M.}~\bibnamefont
  {Laakso}}\ and\ \bibinfo {author} {\bibfnamefont {M.}~\bibnamefont
  {Pletyukhov}},\ }\href {\doibase 10.1103/PhysRevLett.113.183601} {\bibfield
  {journal} {\bibinfo  {journal} {Phys. Rev. Lett.}\ }\textbf {\bibinfo
  {volume} {113}},\ \bibinfo {pages} {183601} (\bibinfo {year}
  {2014})}\BibitemShut {NoStop}%
\bibitem [{\citenamefont {Yang}\ \emph {et~al.}(2020)\citenamefont {Yang},
  \citenamefont {Jia},\ and\ \citenamefont
  {Yuan}}]{Yang-Ann.Phys.(Berlin)2020}%
  \BibitemOpen
  \bibfield  {author} {\bibinfo {author} {\bibfnamefont {S.~Y.}\ \bibnamefont
  {Yang}}, \bibinfo {author} {\bibfnamefont {W.~Z.}\ \bibnamefont {Jia}}, \
  and\ \bibinfo {author} {\bibfnamefont {H.}~\bibnamefont {Yuan}},\ }\href
  {\doibase https://doi.org/10.1002/andp.202000154} {\bibfield  {journal}
  {\bibinfo  {journal} {Ann. Phys.}\ }\textbf {\bibinfo {volume} {532}},\
  \bibinfo {pages} {2000154} (\bibinfo {year} {2020})}\BibitemShut {NoStop}%
\bibitem [{\citenamefont {Abdumalikov}\ \emph {et~al.}(2010)\citenamefont
  {Abdumalikov}, \citenamefont {Astafiev}, \citenamefont {Zagoskin},
  \citenamefont {Pashkin}, \citenamefont {Nakamura},\ and\ \citenamefont
  {Tsai}}]{Abdumalikov-PRL2010}%
  \BibitemOpen
  \bibfield  {author} {\bibinfo {author} {\bibfnamefont {A.~A.}\ \bibnamefont
  {Abdumalikov}}, \bibinfo {author} {\bibfnamefont {O.}~\bibnamefont
  {Astafiev}}, \bibinfo {author} {\bibfnamefont {A.~M.}\ \bibnamefont
  {Zagoskin}}, \bibinfo {author} {\bibfnamefont {Y.~A.}\ \bibnamefont
  {Pashkin}}, \bibinfo {author} {\bibfnamefont {Y.}~\bibnamefont {Nakamura}}, \
  and\ \bibinfo {author} {\bibfnamefont {J.~S.}\ \bibnamefont {Tsai}},\ }\href
  {\doibase 10.1103/PhysRevLett.104.193601} {\bibfield  {journal} {\bibinfo
  {journal} {Phys. Rev. Lett.}\ }\textbf {\bibinfo {volume} {104}},\ \bibinfo
  {pages} {193601} (\bibinfo {year} {2010})}\BibitemShut {NoStop}%
\bibitem [{\citenamefont {Bermel}\ \emph {et~al.}(2006)\citenamefont {Bermel},
  \citenamefont {Rodriguez}, \citenamefont {Johnson}, \citenamefont
  {Joannopoulos},\ and\ \citenamefont {Solja\ifmmode \check{c}\else
  \v{c}\fi{}i\ifmmode~\acute{c}\else \'{c}\fi{}}}]{Bermel-PRA2006}%
  \BibitemOpen
  \bibfield  {author} {\bibinfo {author} {\bibfnamefont {P.}~\bibnamefont
  {Bermel}}, \bibinfo {author} {\bibfnamefont {A.}~\bibnamefont {Rodriguez}},
  \bibinfo {author} {\bibfnamefont {S.~G.}\ \bibnamefont {Johnson}}, \bibinfo
  {author} {\bibfnamefont {J.~D.}\ \bibnamefont {Joannopoulos}}, \ and\
  \bibinfo {author} {\bibfnamefont {M.}~\bibnamefont {Solja\ifmmode
  \check{c}\else \v{c}\fi{}i\ifmmode~\acute{c}\else \'{c}\fi{}}},\ }\href
  {\doibase 10.1103/PhysRevA.74.043818} {\bibfield  {journal} {\bibinfo
  {journal} {Phys. Rev. A}\ }\textbf {\bibinfo {volume} {74}},\ \bibinfo
  {pages} {043818} (\bibinfo {year} {2006})}\BibitemShut {NoStop}%
\bibitem [{\citenamefont {Aoki}\ \emph {et~al.}(2009)\citenamefont {Aoki},
  \citenamefont {Parkins}, \citenamefont {Alton}, \citenamefont {Regal},
  \citenamefont {Dayan}, \citenamefont {Ostby}, \citenamefont {Vahala},\ and\
  \citenamefont {Kimble}}]{Aoki-PRL2009}%
  \BibitemOpen
  \bibfield  {author} {\bibinfo {author} {\bibfnamefont {T.}~\bibnamefont
  {Aoki}}, \bibinfo {author} {\bibfnamefont {A.~S.}\ \bibnamefont {Parkins}},
  \bibinfo {author} {\bibfnamefont {D.~J.}\ \bibnamefont {Alton}}, \bibinfo
  {author} {\bibfnamefont {C.~A.}\ \bibnamefont {Regal}}, \bibinfo {author}
  {\bibfnamefont {B.}~\bibnamefont {Dayan}}, \bibinfo {author} {\bibfnamefont
  {E.}~\bibnamefont {Ostby}}, \bibinfo {author} {\bibfnamefont {K.~J.}\
  \bibnamefont {Vahala}}, \ and\ \bibinfo {author} {\bibfnamefont {H.~J.}\
  \bibnamefont {Kimble}},\ }\href {\doibase 10.1103/PhysRevLett.102.083601}
  {\bibfield  {journal} {\bibinfo  {journal} {Phys. Rev. Lett.}\ }\textbf
  {\bibinfo {volume} {102}},\ \bibinfo {pages} {083601} (\bibinfo {year}
  {2009})}\BibitemShut {NoStop}%
\bibitem [{\citenamefont {Zhu}\ and\ \citenamefont {Jia}(2019)}]{Zhu-PRA2019}%
  \BibitemOpen
  \bibfield  {author} {\bibinfo {author} {\bibfnamefont {Y.~T.}\ \bibnamefont
  {Zhu}}\ and\ \bibinfo {author} {\bibfnamefont {W.~Z.}\ \bibnamefont {Jia}},\
  }\href {\doibase 10.1103/PhysRevA.99.063815} {\bibfield  {journal} {\bibinfo
  {journal} {Phys. Rev. A}\ }\textbf {\bibinfo {volume} {99}},\ \bibinfo
  {pages} {063815} (\bibinfo {year} {2019})}\BibitemShut {NoStop}%
\bibitem [{\citenamefont {Neumeier}\ \emph {et~al.}(2013)\citenamefont
  {Neumeier}, \citenamefont {Leib},\ and\ \citenamefont
  {Hartmann}}]{Neumeier-PRL2013}%
  \BibitemOpen
  \bibfield  {author} {\bibinfo {author} {\bibfnamefont {L.}~\bibnamefont
  {Neumeier}}, \bibinfo {author} {\bibfnamefont {M.}~\bibnamefont {Leib}}, \
  and\ \bibinfo {author} {\bibfnamefont {M.~J.}\ \bibnamefont {Hartmann}},\
  }\href {\doibase 10.1103/PhysRevLett.111.063601} {\bibfield  {journal}
  {\bibinfo  {journal} {Phys. Rev. Lett.}\ }\textbf {\bibinfo {volume} {111}},\
  \bibinfo {pages} {063601} (\bibinfo {year} {2013})}\BibitemShut {NoStop}%
\bibitem [{\citenamefont {Bradford}\ \emph {et~al.}(2012)\citenamefont
  {Bradford}, \citenamefont {Obi},\ and\ \citenamefont
  {Shen}}]{Bradford-PRL2012}%
  \BibitemOpen
  \bibfield  {author} {\bibinfo {author} {\bibfnamefont {M.}~\bibnamefont
  {Bradford}}, \bibinfo {author} {\bibfnamefont {K.~C.}\ \bibnamefont {Obi}}, \
  and\ \bibinfo {author} {\bibfnamefont {J.-T.}\ \bibnamefont {Shen}},\ }\href
  {\doibase 10.1103/PhysRevLett.108.103902} {\bibfield  {journal} {\bibinfo
  {journal} {Phys. Rev. Lett.}\ }\textbf {\bibinfo {volume} {108}},\ \bibinfo
  {pages} {103902} (\bibinfo {year} {2012})}\BibitemShut {NoStop}%
\bibitem [{\citenamefont {Bradford}\ and\ \citenamefont
  {Shen}(2012)}]{Bradford-PRA2012}%
  \BibitemOpen
  \bibfield  {author} {\bibinfo {author} {\bibfnamefont {M.}~\bibnamefont
  {Bradford}}\ and\ \bibinfo {author} {\bibfnamefont {J.-T.}\ \bibnamefont
  {Shen}},\ }\href {\doibase 10.1103/PhysRevA.85.043814} {\bibfield  {journal}
  {\bibinfo  {journal} {Phys. Rev. A}\ }\textbf {\bibinfo {volume} {85}},\
  \bibinfo {pages} {043814} (\bibinfo {year} {2012})}\BibitemShut {NoStop}%
\bibitem [{\citenamefont {Wang}\ \emph {et~al.}(2014)\citenamefont {Wang},
  \citenamefont {Zhou}, \citenamefont {Li},\ and\ \citenamefont
  {Sun}}]{Wang-PRA2014}%
  \BibitemOpen
  \bibfield  {author} {\bibinfo {author} {\bibfnamefont {Z.~H.}\ \bibnamefont
  {Wang}}, \bibinfo {author} {\bibfnamefont {L.}~\bibnamefont {Zhou}}, \bibinfo
  {author} {\bibfnamefont {Y.}~\bibnamefont {Li}}, \ and\ \bibinfo {author}
  {\bibfnamefont {C.~P.}\ \bibnamefont {Sun}},\ }\href {\doibase
  10.1103/PhysRevA.89.053813} {\bibfield  {journal} {\bibinfo  {journal} {Phys.
  Rev. A}\ }\textbf {\bibinfo {volume} {89}},\ \bibinfo {pages} {053813}
  (\bibinfo {year} {2014})}\BibitemShut {NoStop}%
\bibitem [{\citenamefont {Zhao}\ \emph {et~al.}(2017)\citenamefont {Zhao},
  \citenamefont {Ding}, \citenamefont {Peng},\ and\ \citenamefont
  {Liu}}]{Zhao-PRA2017}%
  \BibitemOpen
  \bibfield  {author} {\bibinfo {author} {\bibfnamefont {Y.-J.}\ \bibnamefont
  {Zhao}}, \bibinfo {author} {\bibfnamefont {J.-H.}\ \bibnamefont {Ding}},
  \bibinfo {author} {\bibfnamefont {Z.~H.}\ \bibnamefont {Peng}}, \ and\
  \bibinfo {author} {\bibfnamefont {Y.-x.}\ \bibnamefont {Liu}},\ }\href
  {\doibase 10.1103/PhysRevA.95.043806} {\bibfield  {journal} {\bibinfo
  {journal} {Phys. Rev. A}\ }\textbf {\bibinfo {volume} {95}},\ \bibinfo
  {pages} {043806} (\bibinfo {year} {2017})}\BibitemShut {NoStop}%
\bibitem [{\citenamefont {Jia}\ \emph {et~al.}(2017)\citenamefont {Jia},
  \citenamefont {Wang},\ and\ \citenamefont {Liu}}]{Jia-PRA2017}%
  \BibitemOpen
  \bibfield  {author} {\bibinfo {author} {\bibfnamefont {W.~Z.}\ \bibnamefont
  {Jia}}, \bibinfo {author} {\bibfnamefont {Y.~W.}\ \bibnamefont {Wang}}, \
  and\ \bibinfo {author} {\bibfnamefont {Y.-x.}\ \bibnamefont {Liu}},\ }\href
  {\doibase 10.1103/PhysRevA.96.053832} {\bibfield  {journal} {\bibinfo
  {journal} {Phys. Rev. A}\ }\textbf {\bibinfo {volume} {96}},\ \bibinfo
  {pages} {053832} (\bibinfo {year} {2017})}\BibitemShut {NoStop}%
\bibitem [{\citenamefont {Cao}\ and\ \citenamefont {Jia}(2021)}]{Cao-JPB2021}%
  \BibitemOpen
  \bibfield  {author} {\bibinfo {author} {\bibfnamefont {M.~S.}\ \bibnamefont
  {Cao}}\ and\ \bibinfo {author} {\bibfnamefont {W.~Z.}\ \bibnamefont {Jia}},\
  }\href {\doibase 10.1088/1361-6455/abe395} {\bibfield  {journal} {\bibinfo
  {journal} {J. Phys. B}\ }\textbf {\bibinfo {volume} {54}},\ \bibinfo {pages}
  {055502} (\bibinfo {year} {2021})}\BibitemShut {NoStop}%
\bibitem [{\citenamefont {Tsoi}\ and\ \citenamefont
  {Law}(2008)}]{Tsoi-PRA2008}%
  \BibitemOpen
  \bibfield  {author} {\bibinfo {author} {\bibfnamefont {T.~S.}\ \bibnamefont
  {Tsoi}}\ and\ \bibinfo {author} {\bibfnamefont {C.~K.}\ \bibnamefont {Law}},\
  }\href {\doibase 10.1103/PhysRevA.78.063832} {\bibfield  {journal} {\bibinfo
  {journal} {Phys. Rev. A}\ }\textbf {\bibinfo {volume} {78}},\ \bibinfo
  {pages} {063832} (\bibinfo {year} {2008})}\BibitemShut {NoStop}%
\bibitem [{\citenamefont {Cheng}\ and\ \citenamefont
  {Song}(2012)}]{Cheng-OSA2012}%
  \BibitemOpen
  \bibfield  {author} {\bibinfo {author} {\bibfnamefont {M.-T.}\ \bibnamefont
  {Cheng}}\ and\ \bibinfo {author} {\bibfnamefont {Y.-Y.}\ \bibnamefont
  {Song}},\ }\href {\doibase 10.1364/OL.37.000978} {\bibfield  {journal}
  {\bibinfo  {journal} {Opt. Lett.}\ }\textbf {\bibinfo {volume} {37}},\
  \bibinfo {pages} {978} (\bibinfo {year} {2012})}\BibitemShut {NoStop}%
\bibitem [{\citenamefont {Liao}\ \emph {et~al.}(2015)\citenamefont {Liao},
  \citenamefont {Zeng}, \citenamefont {Zhu},\ and\ \citenamefont
  {Zubairy}}]{Liao-PRA2015}%
  \BibitemOpen
  \bibfield  {author} {\bibinfo {author} {\bibfnamefont {Z.}~\bibnamefont
  {Liao}}, \bibinfo {author} {\bibfnamefont {X.}~\bibnamefont {Zeng}}, \bibinfo
  {author} {\bibfnamefont {S.-Y.}\ \bibnamefont {Zhu}}, \ and\ \bibinfo
  {author} {\bibfnamefont {M.~S.}\ \bibnamefont {Zubairy}},\ }\href {\doibase
  10.1103/PhysRevA.92.023806} {\bibfield  {journal} {\bibinfo  {journal} {Phys.
  Rev. A}\ }\textbf {\bibinfo {volume} {92}},\ \bibinfo {pages} {023806}
  (\bibinfo {year} {2015})}\BibitemShut {NoStop}%
\bibitem [{\citenamefont {Cheng}\ \emph {et~al.}(2017)\citenamefont {Cheng},
  \citenamefont {Xu},\ and\ \citenamefont {Agarwal}}]{Cheng-PRA2017}%
  \BibitemOpen
  \bibfield  {author} {\bibinfo {author} {\bibfnamefont {M.-T.}\ \bibnamefont
  {Cheng}}, \bibinfo {author} {\bibfnamefont {J.}~\bibnamefont {Xu}}, \ and\
  \bibinfo {author} {\bibfnamefont {G.~S.}\ \bibnamefont {Agarwal}},\ }\href
  {\doibase 10.1103/PhysRevA.95.053807} {\bibfield  {journal} {\bibinfo
  {journal} {Phys. Rev. A}\ }\textbf {\bibinfo {volume} {95}},\ \bibinfo
  {pages} {053807} (\bibinfo {year} {2017})}\BibitemShut {NoStop}%
\bibitem [{\citenamefont {Mukhopadhyay}\ and\ \citenamefont
  {Agarwal}(2019)}]{Mukhopadhyay-PRA2019}%
  \BibitemOpen
  \bibfield  {author} {\bibinfo {author} {\bibfnamefont {D.}~\bibnamefont
  {Mukhopadhyay}}\ and\ \bibinfo {author} {\bibfnamefont {G.~S.}\ \bibnamefont
  {Agarwal}},\ }\href {\doibase 10.1103/PhysRevA.100.013812} {\bibfield
  {journal} {\bibinfo  {journal} {Phys. Rev. A}\ }\textbf {\bibinfo {volume}
  {100}},\ \bibinfo {pages} {013812} (\bibinfo {year} {2019})}\BibitemShut
  {NoStop}%
\bibitem [{\citenamefont {Shen}\ \emph {et~al.}(2007)\citenamefont {Shen},
  \citenamefont {Povinelli}, \citenamefont {Sandhu},\ and\ \citenamefont
  {Fan}}]{Shen-PRB2007}%
  \BibitemOpen
  \bibfield  {author} {\bibinfo {author} {\bibfnamefont {J.-T.}\ \bibnamefont
  {Shen}}, \bibinfo {author} {\bibfnamefont {M.~L.}\ \bibnamefont {Povinelli}},
  \bibinfo {author} {\bibfnamefont {S.}~\bibnamefont {Sandhu}}, \ and\ \bibinfo
  {author} {\bibfnamefont {S.}~\bibnamefont {Fan}},\ }\href {\doibase
  10.1103/PhysRevB.75.035320} {\bibfield  {journal} {\bibinfo  {journal} {Phys.
  Rev. B}\ }\textbf {\bibinfo {volume} {75}},\ \bibinfo {pages} {035320}
  (\bibinfo {year} {2007})}\BibitemShut {NoStop}%
\bibitem [{\citenamefont {Fang}\ and\ \citenamefont
  {Baranger}(2017)}]{Fang-PRA2017}%
  \BibitemOpen
  \bibfield  {author} {\bibinfo {author} {\bibfnamefont {Y.-L.~L.}\
  \bibnamefont {Fang}}\ and\ \bibinfo {author} {\bibfnamefont {H.~U.}\
  \bibnamefont {Baranger}},\ }\href {\doibase 10.1103/PhysRevA.96.013842}
  {\bibfield  {journal} {\bibinfo  {journal} {Phys. Rev. A}\ }\textbf {\bibinfo
  {volume} {96}},\ \bibinfo {pages} {013842} (\bibinfo {year}
  {2017})}\BibitemShut {NoStop}%
\bibitem [{\citenamefont {Mukhopadhyay}\ and\ \citenamefont
  {Agarwal}(2020)}]{Mukhopadhyay-PRA2020}%
  \BibitemOpen
  \bibfield  {author} {\bibinfo {author} {\bibfnamefont {D.}~\bibnamefont
  {Mukhopadhyay}}\ and\ \bibinfo {author} {\bibfnamefont {G.~S.}\ \bibnamefont
  {Agarwal}},\ }\href {\doibase 10.1103/PhysRevA.101.063814} {\bibfield
  {journal} {\bibinfo  {journal} {Phys. Rev. A}\ }\textbf {\bibinfo {volume}
  {101}},\ \bibinfo {pages} {063814} (\bibinfo {year} {2020})}\BibitemShut
  {NoStop}%
\bibitem [{\citenamefont {Zheng}\ and\ \citenamefont
  {Baranger}(2013)}]{Zheng-PRL2013}%
  \BibitemOpen
  \bibfield  {author} {\bibinfo {author} {\bibfnamefont {H.}~\bibnamefont
  {Zheng}}\ and\ \bibinfo {author} {\bibfnamefont {H.~U.}\ \bibnamefont
  {Baranger}},\ }\href {\doibase 10.1103/PhysRevLett.110.113601} {\bibfield
  {journal} {\bibinfo  {journal} {Phys. Rev. Lett.}\ }\textbf {\bibinfo
  {volume} {110}},\ \bibinfo {pages} {113601} (\bibinfo {year}
  {2013})}\BibitemShut {NoStop}%
\bibitem [{\citenamefont {Gonzalez-Ballestero}\ \emph
  {et~al.}(2014)\citenamefont {Gonzalez-Ballestero}, \citenamefont {Moreno},\
  and\ \citenamefont {Garcia-Vidal}}]{Ballestero-PRA2014}%
  \BibitemOpen
  \bibfield  {author} {\bibinfo {author} {\bibfnamefont {C.}~\bibnamefont
  {Gonzalez-Ballestero}}, \bibinfo {author} {\bibfnamefont {E.}~\bibnamefont
  {Moreno}}, \ and\ \bibinfo {author} {\bibfnamefont {F.~J.}\ \bibnamefont
  {Garcia-Vidal}},\ }\href {\doibase 10.1103/PhysRevA.89.042328} {\bibfield
  {journal} {\bibinfo  {journal} {Phys. Rev. A}\ }\textbf {\bibinfo {volume}
  {89}},\ \bibinfo {pages} {042328} (\bibinfo {year} {2014})}\BibitemShut
  {NoStop}%
\bibitem [{\citenamefont {Facchi}\ \emph {et~al.}(2016)\citenamefont {Facchi},
  \citenamefont {Kim}, \citenamefont {Pascazio}, \citenamefont {Pepe},
  \citenamefont {Pomarico},\ and\ \citenamefont {Tufarelli}}]{Facchi-PRA2016}%
  \BibitemOpen
  \bibfield  {author} {\bibinfo {author} {\bibfnamefont {P.}~\bibnamefont
  {Facchi}}, \bibinfo {author} {\bibfnamefont {M.~S.}\ \bibnamefont {Kim}},
  \bibinfo {author} {\bibfnamefont {S.}~\bibnamefont {Pascazio}}, \bibinfo
  {author} {\bibfnamefont {F.~V.}\ \bibnamefont {Pepe}}, \bibinfo {author}
  {\bibfnamefont {D.}~\bibnamefont {Pomarico}}, \ and\ \bibinfo {author}
  {\bibfnamefont {T.}~\bibnamefont {Tufarelli}},\ }\href {\doibase
  10.1103/PhysRevA.94.043839} {\bibfield  {journal} {\bibinfo  {journal} {Phys.
  Rev. A}\ }\textbf {\bibinfo {volume} {94}},\ \bibinfo {pages} {043839}
  (\bibinfo {year} {2016})}\BibitemShut {NoStop}%
\bibitem [{\citenamefont {Mirza}\ and\ \citenamefont
  {Schotland}(2016)}]{Mirza-PRA2016}%
  \BibitemOpen
  \bibfield  {author} {\bibinfo {author} {\bibfnamefont {I.~M.}\ \bibnamefont
  {Mirza}}\ and\ \bibinfo {author} {\bibfnamefont {J.~C.}\ \bibnamefont
  {Schotland}},\ }\href {\doibase 10.1103/PhysRevA.94.012302} {\bibfield
  {journal} {\bibinfo  {journal} {Phys. Rev. A}\ }\textbf {\bibinfo {volume}
  {94}},\ \bibinfo {pages} {012302} (\bibinfo {year} {2016})}\BibitemShut
  {NoStop}%
\bibitem [{\citenamefont {Fang}\ and\ \citenamefont
  {Baranger}(2015)}]{Fang-PRA2015}%
  \BibitemOpen
  \bibfield  {author} {\bibinfo {author} {\bibfnamefont {Y.-L.~L.}\
  \bibnamefont {Fang}}\ and\ \bibinfo {author} {\bibfnamefont {H.~U.}\
  \bibnamefont {Baranger}},\ }\href {\doibase 10.1103/PhysRevA.91.053845}
  {\bibfield  {journal} {\bibinfo  {journal} {Phys. Rev. A}\ }\textbf {\bibinfo
  {volume} {91}},\ \bibinfo {pages} {053845} (\bibinfo {year}
  {2015})}\BibitemShut {NoStop}%
\bibitem [{\citenamefont {Greenberg}\ \emph {et~al.}(2021)\citenamefont
  {Greenberg}, \citenamefont {Shtygashev},\ and\ \citenamefont
  {Moiseev}}]{Greenberg-PRA2021}%
  \BibitemOpen
  \bibfield  {author} {\bibinfo {author} {\bibfnamefont {Y.~S.}\ \bibnamefont
  {Greenberg}}, \bibinfo {author} {\bibfnamefont {A.~A.}\ \bibnamefont
  {Shtygashev}}, \ and\ \bibinfo {author} {\bibfnamefont {A.~G.}\ \bibnamefont
  {Moiseev}},\ }\href {\doibase 10.1103/PhysRevA.103.023508} {\bibfield
  {journal} {\bibinfo  {journal} {Phys. Rev. A}\ }\textbf {\bibinfo {volume}
  {103}},\ \bibinfo {pages} {023508} (\bibinfo {year} {2021})}\BibitemShut
  {NoStop}%
\bibitem [{\citenamefont {Chang}\ \emph {et~al.}(2012)\citenamefont {Chang},
  \citenamefont {Jiang}, \citenamefont {Gorshkov},\ and\ \citenamefont
  {Kimble}}]{Chang-IOP2012}%
  \BibitemOpen
  \bibfield  {author} {\bibinfo {author} {\bibfnamefont {D.~E.}\ \bibnamefont
  {Chang}}, \bibinfo {author} {\bibfnamefont {L.}~\bibnamefont {Jiang}},
  \bibinfo {author} {\bibfnamefont {A.~V.}\ \bibnamefont {Gorshkov}}, \ and\
  \bibinfo {author} {\bibfnamefont {H.~J.}\ \bibnamefont {Kimble}},\ }\href
  {\doibase 10.1088/1367-2630/14/6/063003} {\bibfield  {journal} {\bibinfo
  {journal} {New J. Phys.}\ }\textbf {\bibinfo {volume} {14}},\ \bibinfo
  {pages} {063003} (\bibinfo {year} {2012})}\BibitemShut {NoStop}%
\bibitem [{\citenamefont {Mirhosseini}\ \emph {et~al.}(2019)\citenamefont
  {Mirhosseini}, \citenamefont {Kim}, \citenamefont {Zhang}, \citenamefont
  {Sipahigil}, \citenamefont {Dieterle}, \citenamefont {Keller}, \citenamefont
  {Asenjo-Garcia}, \citenamefont {Chang},\ and\ \citenamefont
  {Painter}}]{Mirhosseini-Nature2019}%
  \BibitemOpen
  \bibfield  {author} {\bibinfo {author} {\bibfnamefont {M.}~\bibnamefont
  {Mirhosseini}}, \bibinfo {author} {\bibfnamefont {E.}~\bibnamefont {Kim}},
  \bibinfo {author} {\bibfnamefont {X.}~\bibnamefont {Zhang}}, \bibinfo
  {author} {\bibfnamefont {A.}~\bibnamefont {Sipahigil}}, \bibinfo {author}
  {\bibfnamefont {P.~B.}\ \bibnamefont {Dieterle}}, \bibinfo {author}
  {\bibfnamefont {A.~J.}\ \bibnamefont {Keller}}, \bibinfo {author}
  {\bibfnamefont {A.}~\bibnamefont {Asenjo-Garcia}}, \bibinfo {author}
  {\bibfnamefont {D.~E.}\ \bibnamefont {Chang}}, \ and\ \bibinfo {author}
  {\bibfnamefont {O.}~\bibnamefont {Painter}},\ }\href {\doibase
  10.1038/s41586-019-1196-1} {\bibfield  {journal} {\bibinfo  {journal}
  {Nature}\ }\textbf {\bibinfo {volume} {569}},\ \bibinfo {pages} {692}
  (\bibinfo {year} {2019})}\BibitemShut {NoStop}%
\bibitem [{\citenamefont {van Loo}\ \emph {et~al.}(2013)\citenamefont {van
  Loo}, \citenamefont {Fedorov}, \citenamefont {Lalumi{\`e}re}, \citenamefont
  {Sanders}, \citenamefont {Blais},\ and\ \citenamefont
  {Wallraff}}]{Loo-Sci2013}%
  \BibitemOpen
  \bibfield  {author} {\bibinfo {author} {\bibfnamefont {A.~F.}\ \bibnamefont
  {van Loo}}, \bibinfo {author} {\bibfnamefont {A.}~\bibnamefont {Fedorov}},
  \bibinfo {author} {\bibfnamefont {K.}~\bibnamefont {Lalumi{\`e}re}}, \bibinfo
  {author} {\bibfnamefont {B.~C.}\ \bibnamefont {Sanders}}, \bibinfo {author}
  {\bibfnamefont {A.}~\bibnamefont {Blais}}, \ and\ \bibinfo {author}
  {\bibfnamefont {A.}~\bibnamefont {Wallraff}},\ }\href {\doibase
  10.1126/science.1244324} {\bibfield  {journal} {\bibinfo  {journal}
  {Science}\ }\textbf {\bibinfo {volume} {342}},\ \bibinfo {pages} {1494}
  (\bibinfo {year} {2013})}\BibitemShut {NoStop}%
\bibitem [{\citenamefont {Zhang}\ and\ \citenamefont
  {M\o{}lmer}(2019)}]{Zhang-PRL2019}%
  \BibitemOpen
  \bibfield  {author} {\bibinfo {author} {\bibfnamefont {Y.-X.}\ \bibnamefont
  {Zhang}}\ and\ \bibinfo {author} {\bibfnamefont {K.}~\bibnamefont
  {M\o{}lmer}},\ }\href {\doibase 10.1103/PhysRevLett.122.203605} {\bibfield
  {journal} {\bibinfo  {journal} {Phys. Rev. Lett.}\ }\textbf {\bibinfo
  {volume} {122}},\ \bibinfo {pages} {203605} (\bibinfo {year}
  {2019})}\BibitemShut {NoStop}%
\bibitem [{\citenamefont {Ke}\ \emph {et~al.}(2019)\citenamefont {Ke},
  \citenamefont {Poshakinskiy}, \citenamefont {Lee}, \citenamefont {Kivshar},\
  and\ \citenamefont {Poddubny}}]{Ke-PRL2019}%
  \BibitemOpen
  \bibfield  {author} {\bibinfo {author} {\bibfnamefont {Y.}~\bibnamefont
  {Ke}}, \bibinfo {author} {\bibfnamefont {A.~V.}\ \bibnamefont
  {Poshakinskiy}}, \bibinfo {author} {\bibfnamefont {C.}~\bibnamefont {Lee}},
  \bibinfo {author} {\bibfnamefont {Y.~S.}\ \bibnamefont {Kivshar}}, \ and\
  \bibinfo {author} {\bibfnamefont {A.~N.}\ \bibnamefont {Poddubny}},\ }\href
  {\doibase 10.1103/PhysRevLett.123.253601} {\bibfield  {journal} {\bibinfo
  {journal} {Phys. Rev. Lett.}\ }\textbf {\bibinfo {volume} {123}},\ \bibinfo
  {pages} {253601} (\bibinfo {year} {2019})}\BibitemShut {NoStop}%
\bibitem [{\citenamefont {Wang}\ \emph {et~al.}(2020)\citenamefont {Wang},
  \citenamefont {Li}, \citenamefont {Feng}, \citenamefont {Song}, \citenamefont
  {Song}, \citenamefont {Liu}, \citenamefont {Guo}, \citenamefont {Zhang},
  \citenamefont {Dong}, \citenamefont {Zheng}, \citenamefont {Wang},\ and\
  \citenamefont {Wang}}]{Wang-PRL2020}%
  \BibitemOpen
  \bibfield  {author} {\bibinfo {author} {\bibfnamefont {Z.}~\bibnamefont
  {Wang}}, \bibinfo {author} {\bibfnamefont {H.}~\bibnamefont {Li}}, \bibinfo
  {author} {\bibfnamefont {W.}~\bibnamefont {Feng}}, \bibinfo {author}
  {\bibfnamefont {X.}~\bibnamefont {Song}}, \bibinfo {author} {\bibfnamefont
  {C.}~\bibnamefont {Song}}, \bibinfo {author} {\bibfnamefont {W.}~\bibnamefont
  {Liu}}, \bibinfo {author} {\bibfnamefont {Q.}~\bibnamefont {Guo}}, \bibinfo
  {author} {\bibfnamefont {X.}~\bibnamefont {Zhang}}, \bibinfo {author}
  {\bibfnamefont {H.}~\bibnamefont {Dong}}, \bibinfo {author} {\bibfnamefont
  {D.}~\bibnamefont {Zheng}}, \bibinfo {author} {\bibfnamefont
  {H.}~\bibnamefont {Wang}}, \ and\ \bibinfo {author} {\bibfnamefont {D.-W.}\
  \bibnamefont {Wang}},\ }\href {\doibase 10.1103/PhysRevLett.124.013601}
  {\bibfield  {journal} {\bibinfo  {journal} {Phys. Rev. Lett.}\ }\textbf
  {\bibinfo {volume} {124}},\ \bibinfo {pages} {013601} (\bibinfo {year}
  {2020})}\BibitemShut {NoStop}%
\bibitem [{\citenamefont {Kockum}(2021)}]{Kockum-MI2020}%
  \BibitemOpen
  \bibfield  {author} {\bibinfo {author} {\bibfnamefont {A.~F.}\ \bibnamefont
  {Kockum}},\ }in\ \href {\doibase 10.1007/978-981-15-5191-8_12} {\emph
  {\bibinfo {booktitle} {International Symposium on Mathematics, Quantum
  Theory, and Cryptography}}},\ \bibinfo {editor} {edited by\ \bibinfo {editor}
  {\bibfnamefont {T.}~\bibnamefont {Takagi}}, \bibinfo {editor} {\bibfnamefont
  {M.}~\bibnamefont {Wakayama}}, \bibinfo {editor} {\bibfnamefont
  {K.}~\bibnamefont {Tanaka}}, \bibinfo {editor} {\bibfnamefont
  {N.}~\bibnamefont {Kunihiro}}, \bibinfo {editor} {\bibfnamefont
  {K.}~\bibnamefont {Kimoto}}, \ and\ \bibinfo {editor} {\bibfnamefont
  {Y.}~\bibnamefont {Ikematsu}}}\ (\bibinfo  {publisher} {Springer Singapore},\
  \bibinfo {address} {Singapore},\ \bibinfo {year} {2021})\ pp.\ \bibinfo
  {pages} {125--146}\BibitemShut {NoStop}%
\bibitem [{\citenamefont {Kockum}\ \emph {et~al.}(2014)\citenamefont {Kockum},
  \citenamefont {Delsing},\ and\ \citenamefont {Johansson}}]{Kockum-PRA2014}%
  \BibitemOpen
  \bibfield  {author} {\bibinfo {author} {\bibfnamefont {A.~F.}\ \bibnamefont
  {Kockum}}, \bibinfo {author} {\bibfnamefont {P.}~\bibnamefont {Delsing}}, \
  and\ \bibinfo {author} {\bibfnamefont {G.}~\bibnamefont {Johansson}},\ }\href
  {\doibase 10.1103/PhysRevA.90.013837} {\bibfield  {journal} {\bibinfo
  {journal} {Phys. Rev. A}\ }\textbf {\bibinfo {volume} {90}},\ \bibinfo
  {pages} {013837} (\bibinfo {year} {2014})}\BibitemShut {NoStop}%
\bibitem [{\citenamefont {Koch}\ \emph {et~al.}(2007)\citenamefont {Koch},
  \citenamefont {Yu}, \citenamefont {Gambetta}, \citenamefont {Houck},
  \citenamefont {Schuster}, \citenamefont {Majer}, \citenamefont {Blais},
  \citenamefont {Devoret}, \citenamefont {Girvin},\ and\ \citenamefont
  {Schoelkopf}}]{Koch-PRA2007}%
  \BibitemOpen
  \bibfield  {author} {\bibinfo {author} {\bibfnamefont {J.}~\bibnamefont
  {Koch}}, \bibinfo {author} {\bibfnamefont {T.~M.}\ \bibnamefont {Yu}},
  \bibinfo {author} {\bibfnamefont {J.}~\bibnamefont {Gambetta}}, \bibinfo
  {author} {\bibfnamefont {A.~A.}\ \bibnamefont {Houck}}, \bibinfo {author}
  {\bibfnamefont {D.~I.}\ \bibnamefont {Schuster}}, \bibinfo {author}
  {\bibfnamefont {J.}~\bibnamefont {Majer}}, \bibinfo {author} {\bibfnamefont
  {A.}~\bibnamefont {Blais}}, \bibinfo {author} {\bibfnamefont {M.~H.}\
  \bibnamefont {Devoret}}, \bibinfo {author} {\bibfnamefont {S.~M.}\
  \bibnamefont {Girvin}}, \ and\ \bibinfo {author} {\bibfnamefont {R.~J.}\
  \bibnamefont {Schoelkopf}},\ }\href {\doibase 10.1103/PhysRevA.76.042319}
  {\bibfield  {journal} {\bibinfo  {journal} {Phys. Rev. A}\ }\textbf {\bibinfo
  {volume} {76}},\ \bibinfo {pages} {042319} (\bibinfo {year}
  {2007})}\BibitemShut {NoStop}%
\bibitem [{\citenamefont {Kannan}\ \emph {et~al.}(2020)\citenamefont {Kannan},
  \citenamefont {Ruckriegel}, \citenamefont {Campbell}, \citenamefont {Kockum},
  \citenamefont {Braumüller}, \citenamefont {Kim}, \citenamefont {Kjaergaard},
  \citenamefont {Krantz}, \citenamefont {Melville}, \citenamefont
  {Niedzielski}, \citenamefont {Vepsäläinen}, \citenamefont {Winik},
  \citenamefont {Yoder}, \citenamefont {Nori}, \citenamefont {Orlando},
  \citenamefont {Gustavsson},\ and\ \citenamefont
  {Oliver}}]{Kannan-Nature2020}%
  \BibitemOpen
  \bibfield  {author} {\bibinfo {author} {\bibfnamefont {B.}~\bibnamefont
  {Kannan}}, \bibinfo {author} {\bibfnamefont {M.~J.}\ \bibnamefont
  {Ruckriegel}}, \bibinfo {author} {\bibfnamefont {D.~L.}\ \bibnamefont
  {Campbell}}, \bibinfo {author} {\bibfnamefont {A.~F.}\ \bibnamefont
  {Kockum}}, \bibinfo {author} {\bibfnamefont {J.}~\bibnamefont {Braumüller}},
  \bibinfo {author} {\bibfnamefont {D.~K.}\ \bibnamefont {Kim}}, \bibinfo
  {author} {\bibfnamefont {M.}~\bibnamefont {Kjaergaard}}, \bibinfo {author}
  {\bibfnamefont {P.}~\bibnamefont {Krantz}}, \bibinfo {author} {\bibfnamefont
  {A.}~\bibnamefont {Melville}}, \bibinfo {author} {\bibfnamefont {B.~M.}\
  \bibnamefont {Niedzielski}}, \bibinfo {author} {\bibfnamefont
  {A.}~\bibnamefont {Vepsäläinen}}, \bibinfo {author} {\bibfnamefont
  {R.}~\bibnamefont {Winik}}, \bibinfo {author} {\bibfnamefont {J.~L.}\
  \bibnamefont {Yoder}}, \bibinfo {author} {\bibfnamefont {F.}~\bibnamefont
  {Nori}}, \bibinfo {author} {\bibfnamefont {T.~P.}\ \bibnamefont {Orlando}},
  \bibinfo {author} {\bibfnamefont {S.}~\bibnamefont {Gustavsson}}, \ and\
  \bibinfo {author} {\bibfnamefont {W.~D.}\ \bibnamefont {Oliver}},\ }\href
  {\doibase 10.1038/s41586-020-2529-9} {\bibfield  {journal} {\bibinfo
  {journal} {Nature}\ }\textbf {\bibinfo {volume} {583}},\ \bibinfo {pages}
  {775} (\bibinfo {year} {2020})}\BibitemShut {NoStop}%
\bibitem [{\citenamefont {Vadiraj}\ \emph {et~al.}(2021)\citenamefont
  {Vadiraj}, \citenamefont {Ask}, \citenamefont {McConkey}, \citenamefont
  {Nsanzineza}, \citenamefont {Chang}, \citenamefont {Kockum},\ and\
  \citenamefont {Wilson}}]{Vadiraj-PRA2021}%
  \BibitemOpen
  \bibfield  {author} {\bibinfo {author} {\bibfnamefont {A.~M.}\ \bibnamefont
  {Vadiraj}}, \bibinfo {author} {\bibfnamefont {A.}~\bibnamefont {Ask}},
  \bibinfo {author} {\bibfnamefont {T.~G.}\ \bibnamefont {McConkey}}, \bibinfo
  {author} {\bibfnamefont {I.}~\bibnamefont {Nsanzineza}}, \bibinfo {author}
  {\bibfnamefont {C.~W.~S.}\ \bibnamefont {Chang}}, \bibinfo {author}
  {\bibfnamefont {A.~F.}\ \bibnamefont {Kockum}}, \ and\ \bibinfo {author}
  {\bibfnamefont {C.~M.}\ \bibnamefont {Wilson}},\ }\href {\doibase
  10.1103/PhysRevA.103.023710} {\bibfield  {journal} {\bibinfo  {journal}
  {Phys. Rev. A}\ }\textbf {\bibinfo {volume} {103}},\ \bibinfo {pages}
  {023710} (\bibinfo {year} {2021})}\BibitemShut {NoStop}%
\bibitem [{\citenamefont {Yu}\ \emph {et~al.}(2021)\citenamefont {Yu},
  \citenamefont {Wang},\ and\ \citenamefont {Wu}}]{Yu-PRA2021}%
  \BibitemOpen
  \bibfield  {author} {\bibinfo {author} {\bibfnamefont {H.}~\bibnamefont
  {Yu}}, \bibinfo {author} {\bibfnamefont {Z.}~\bibnamefont {Wang}}, \ and\
  \bibinfo {author} {\bibfnamefont {J.-H.}\ \bibnamefont {Wu}},\ }\href
  {\doibase 10.1103/PhysRevA.104.013720} {\bibfield  {journal} {\bibinfo
  {journal} {Phys. Rev. A}\ }\textbf {\bibinfo {volume} {104}},\ \bibinfo
  {pages} {013720} (\bibinfo {year} {2021})}\BibitemShut {NoStop}%
\bibitem [{\citenamefont {Kockum}\ \emph {et~al.}(2018)\citenamefont {Kockum},
  \citenamefont {Johansson},\ and\ \citenamefont {Nori}}]{Kockum-PRL2018}%
  \BibitemOpen
  \bibfield  {author} {\bibinfo {author} {\bibfnamefont {A.~F.}\ \bibnamefont
  {Kockum}}, \bibinfo {author} {\bibfnamefont {G.}~\bibnamefont {Johansson}}, \
  and\ \bibinfo {author} {\bibfnamefont {F.}~\bibnamefont {Nori}},\ }\href
  {\doibase 10.1103/PhysRevLett.120.140404} {\bibfield  {journal} {\bibinfo
  {journal} {Phys. Rev. Lett.}\ }\textbf {\bibinfo {volume} {120}},\ \bibinfo
  {pages} {140404} (\bibinfo {year} {2018})}\BibitemShut {NoStop}%
\bibitem [{\citenamefont {Guo}\ \emph {et~al.}(2017)\citenamefont {Guo},
  \citenamefont {Grimsmo}, \citenamefont {Kockum}, \citenamefont {Pletyukhov},\
  and\ \citenamefont {Johansson}}]{Guo-PRA2017}%
  \BibitemOpen
  \bibfield  {author} {\bibinfo {author} {\bibfnamefont {L.}~\bibnamefont
  {Guo}}, \bibinfo {author} {\bibfnamefont {A.}~\bibnamefont {Grimsmo}},
  \bibinfo {author} {\bibfnamefont {A.~F.}\ \bibnamefont {Kockum}}, \bibinfo
  {author} {\bibfnamefont {M.}~\bibnamefont {Pletyukhov}}, \ and\ \bibinfo
  {author} {\bibfnamefont {G.}~\bibnamefont {Johansson}},\ }\href {\doibase
  10.1103/PhysRevA.95.053821} {\bibfield  {journal} {\bibinfo  {journal} {Phys.
  Rev. A}\ }\textbf {\bibinfo {volume} {95}},\ \bibinfo {pages} {053821}
  (\bibinfo {year} {2017})}\BibitemShut {NoStop}%
\bibitem [{\citenamefont {Andersson}\ \emph {et~al.}(2019)\citenamefont
  {Andersson}, \citenamefont {Suri}, \citenamefont {Guo}, \citenamefont
  {Aref},\ and\ \citenamefont {Delsing}}]{Andersson-Nature2019}%
  \BibitemOpen
  \bibfield  {author} {\bibinfo {author} {\bibfnamefont {G.}~\bibnamefont
  {Andersson}}, \bibinfo {author} {\bibfnamefont {B.}~\bibnamefont {Suri}},
  \bibinfo {author} {\bibfnamefont {L.}~\bibnamefont {Guo}}, \bibinfo {author}
  {\bibfnamefont {T.}~\bibnamefont {Aref}}, \ and\ \bibinfo {author}
  {\bibfnamefont {P.}~\bibnamefont {Delsing}},\ }\href {\doibase
  10.1038/s41567-019-0605-6} {\bibfield  {journal} {\bibinfo  {journal} {Nat.
  Phys.}\ }\textbf {\bibinfo {volume} {15}},\ \bibinfo {pages} {1123} (\bibinfo
  {year} {2019})}\BibitemShut {NoStop}%
\bibitem [{\citenamefont {Guo}\ \emph {et~al.}(2020{\natexlab{a}})\citenamefont
  {Guo}, \citenamefont {Kockum}, \citenamefont {Marquardt},\ and\ \citenamefont
  {Johansson}}]{Guo-Phys.Rev.Research2020}%
  \BibitemOpen
  \bibfield  {author} {\bibinfo {author} {\bibfnamefont {L.}~\bibnamefont
  {Guo}}, \bibinfo {author} {\bibfnamefont {A.~F.}\ \bibnamefont {Kockum}},
  \bibinfo {author} {\bibfnamefont {F.}~\bibnamefont {Marquardt}}, \ and\
  \bibinfo {author} {\bibfnamefont {G.}~\bibnamefont {Johansson}},\ }\href
  {\doibase 10.1103/PhysRevResearch.2.043014} {\bibfield  {journal} {\bibinfo
  {journal} {Phys. Rev. Research}\ }\textbf {\bibinfo {volume} {2}},\ \bibinfo
  {pages} {043014} (\bibinfo {year} {2020}{\natexlab{a}})}\BibitemShut
  {NoStop}%
\bibitem [{\citenamefont {Zhao}\ and\ \citenamefont
  {Wang}(2020)}]{Zhao-PRA2020}%
  \BibitemOpen
  \bibfield  {author} {\bibinfo {author} {\bibfnamefont {W.}~\bibnamefont
  {Zhao}}\ and\ \bibinfo {author} {\bibfnamefont {Z.}~\bibnamefont {Wang}},\
  }\href {\doibase 10.1103/PhysRevA.101.053855} {\bibfield  {journal} {\bibinfo
   {journal} {Phys. Rev. A}\ }\textbf {\bibinfo {volume} {101}},\ \bibinfo
  {pages} {053855} (\bibinfo {year} {2020})}\BibitemShut {NoStop}%
\bibitem [{\citenamefont {Guo}\ \emph {et~al.}(2020{\natexlab{b}})\citenamefont
  {Guo}, \citenamefont {Wang}, \citenamefont {Purdy},\ and\ \citenamefont
  {Taylor}}]{Guo-PRA2020}%
  \BibitemOpen
  \bibfield  {author} {\bibinfo {author} {\bibfnamefont {S.}~\bibnamefont
  {Guo}}, \bibinfo {author} {\bibfnamefont {Y.}~\bibnamefont {Wang}}, \bibinfo
  {author} {\bibfnamefont {T.}~\bibnamefont {Purdy}}, \ and\ \bibinfo {author}
  {\bibfnamefont {J.}~\bibnamefont {Taylor}},\ }\href {\doibase
  10.1103/PhysRevA.102.033706} {\bibfield  {journal} {\bibinfo  {journal}
  {Phys. Rev. A}\ }\textbf {\bibinfo {volume} {102}},\ \bibinfo {pages}
  {033706} (\bibinfo {year} {2020}{\natexlab{b}})}\BibitemShut {NoStop}%
\bibitem [{\citenamefont {Ask}\ \emph {et~al.}()\citenamefont {Ask},
  \citenamefont {Fang},\ and\ \citenamefont {Kockum}}]{Ask-arXiv2020}%
  \BibitemOpen
  \bibfield  {author} {\bibinfo {author} {\bibfnamefont {A.}~\bibnamefont
  {Ask}}, \bibinfo {author} {\bibfnamefont {Y.-L.~L.}\ \bibnamefont {Fang}}, \
  and\ \bibinfo {author} {\bibfnamefont {A.~F.}\ \bibnamefont {Kockum}},\
  }\href@noop {} {}\Eprint {http://arxiv.org/abs/2011.15077} {arXiv:2011.15077}
  \BibitemShut {NoStop}%
\bibitem [{\citenamefont {Harris}\ \emph {et~al.}(1990)\citenamefont {Harris},
  \citenamefont {Field},\ and\ \citenamefont {Imamo\ifmmode~\breve{g}\else
  \u{g}\fi{}lu}}]{Harris-PRL1990}%
  \BibitemOpen
  \bibfield  {author} {\bibinfo {author} {\bibfnamefont {S.~E.}\ \bibnamefont
  {Harris}}, \bibinfo {author} {\bibfnamefont {J.~E.}\ \bibnamefont {Field}}, \
  and\ \bibinfo {author} {\bibfnamefont {A.}~\bibnamefont
  {Imamo\ifmmode~\breve{g}\else \u{g}\fi{}lu}},\ }\href {\doibase
  10.1103/PhysRevLett.64.1107} {\bibfield  {journal} {\bibinfo  {journal}
  {Phys. Rev. Lett.}\ }\textbf {\bibinfo {volume} {64}},\ \bibinfo {pages}
  {1107} (\bibinfo {year} {1990})}\BibitemShut {NoStop}%
\bibitem [{\citenamefont {Harris}(1997)}]{Harris-PhysicsToday-1997}%
  \BibitemOpen
  \bibfield  {author} {\bibinfo {author} {\bibfnamefont {S.~E.}\ \bibnamefont
  {Harris}},\ }\href {\doibase 10.1063/1.881806} {\bibfield  {journal}
  {\bibinfo  {journal} {Physics Today}\ }\textbf {\bibinfo {volume} {50}},\
  \bibinfo {pages} {36} (\bibinfo {year} {1997})}\BibitemShut {NoStop}%
\bibitem [{\citenamefont {Fleischhauer}\ \emph {et~al.}(2005)\citenamefont
  {Fleischhauer}, \citenamefont {Imamoglu},\ and\ \citenamefont
  {Marangos}}]{Fleischhauer-Rev.Mod.Phys.2005}%
  \BibitemOpen
  \bibfield  {author} {\bibinfo {author} {\bibfnamefont {M.}~\bibnamefont
  {Fleischhauer}}, \bibinfo {author} {\bibfnamefont {A.}~\bibnamefont
  {Imamoglu}}, \ and\ \bibinfo {author} {\bibfnamefont {J.~P.}\ \bibnamefont
  {Marangos}},\ }\href {\doibase 10.1103/RevModPhys.77.633} {\bibfield
  {journal} {\bibinfo  {journal} {Rev. Mod. Phys.}\ }\textbf {\bibinfo {volume}
  {77}},\ \bibinfo {pages} {633} (\bibinfo {year} {2005})}\BibitemShut
  {NoStop}%
\bibitem [{\citenamefont {Boller}\ \emph {et~al.}(1991)\citenamefont {Boller},
  \citenamefont {Imamo\ifmmode~\breve{g}\else \u{g}\fi{}lu},\ and\
  \citenamefont {Harris}}]{Boller-PRL1991}%
  \BibitemOpen
  \bibfield  {author} {\bibinfo {author} {\bibfnamefont {K.-J.}\ \bibnamefont
  {Boller}}, \bibinfo {author} {\bibfnamefont {A.}~\bibnamefont
  {Imamo\ifmmode~\breve{g}\else \u{g}\fi{}lu}}, \ and\ \bibinfo {author}
  {\bibfnamefont {S.~E.}\ \bibnamefont {Harris}},\ }\href {\doibase
  10.1103/PhysRevLett.66.2593} {\bibfield  {journal} {\bibinfo  {journal}
  {Phys. Rev. Lett.}\ }\textbf {\bibinfo {volume} {66}},\ \bibinfo {pages}
  {2593} (\bibinfo {year} {1991})}\BibitemShut {NoStop}%
\bibitem [{\citenamefont {Fano}(1961)}]{Fano-Phys.Rev1961}%
  \BibitemOpen
  \bibfield  {author} {\bibinfo {author} {\bibfnamefont {U.}~\bibnamefont
  {Fano}},\ }\href {\doibase 10.1103/PhysRev.124.1866} {\bibfield  {journal}
  {\bibinfo  {journal} {Phys. Rev.}\ }\textbf {\bibinfo {volume} {124}},\
  \bibinfo {pages} {1866} (\bibinfo {year} {1961})}\BibitemShut {NoStop}%
\bibitem [{\citenamefont {Miroshnichenko}\ \emph {et~al.}(2010)\citenamefont
  {Miroshnichenko}, \citenamefont {Flach},\ and\ \citenamefont
  {Kivshar}}]{Miroshnichenko-Rev.Mod.Phys.2010}%
  \BibitemOpen
  \bibfield  {author} {\bibinfo {author} {\bibfnamefont {A.~E.}\ \bibnamefont
  {Miroshnichenko}}, \bibinfo {author} {\bibfnamefont {S.}~\bibnamefont
  {Flach}}, \ and\ \bibinfo {author} {\bibfnamefont {Y.~S.}\ \bibnamefont
  {Kivshar}},\ }\href {\doibase 10.1103/RevModPhys.82.2257} {\bibfield
  {journal} {\bibinfo  {journal} {Rev. Mod. Phys.}\ }\textbf {\bibinfo {volume}
  {82}},\ \bibinfo {pages} {2257} (\bibinfo {year} {2010})}\BibitemShut
  {NoStop}%
\bibitem [{\citenamefont {Abi-Salloum}(2010)}]{Abi-Salloum-Phys.Rev.A2010}%
  \BibitemOpen
  \bibfield  {author} {\bibinfo {author} {\bibfnamefont {T.~Y.}\ \bibnamefont
  {Abi-Salloum}},\ }\href {\doibase 10.1103/PhysRevA.81.053836} {\bibfield
  {journal} {\bibinfo  {journal} {Phys. Rev. A}\ }\textbf {\bibinfo {volume}
  {81}},\ \bibinfo {pages} {053836} (\bibinfo {year} {2010})}\BibitemShut
  {NoStop}%
\bibitem [{\citenamefont {Anisimov}\ \emph {et~al.}(2011)\citenamefont
  {Anisimov}, \citenamefont {Dowling},\ and\ \citenamefont
  {Sanders}}]{Anisimov-Phys.Rev.Lett.2011}%
  \BibitemOpen
  \bibfield  {author} {\bibinfo {author} {\bibfnamefont {P.~M.}\ \bibnamefont
  {Anisimov}}, \bibinfo {author} {\bibfnamefont {J.~P.}\ \bibnamefont
  {Dowling}}, \ and\ \bibinfo {author} {\bibfnamefont {B.~C.}\ \bibnamefont
  {Sanders}},\ }\href {\doibase 10.1103/PhysRevLett.107.163604} {\bibfield
  {journal} {\bibinfo  {journal} {Phys. Rev. Lett.}\ }\textbf {\bibinfo
  {volume} {107}},\ \bibinfo {pages} {163604} (\bibinfo {year}
  {2011})}\BibitemShut {NoStop}%
\bibitem [{\citenamefont {Wen}\ \emph {et~al.}(2019)\citenamefont {Wen},
  \citenamefont {Lin}, \citenamefont {Kockum}, \citenamefont {Suri},
  \citenamefont {Ian}, \citenamefont {Chen}, \citenamefont {Mao}, \citenamefont
  {Chiu}, \citenamefont {Delsing}, \citenamefont {Nori}, \citenamefont {Lin},\
  and\ \citenamefont {Hoi}}]{Wen-PhysRevLett.2019}%
  \BibitemOpen
  \bibfield  {author} {\bibinfo {author} {\bibfnamefont {P.~Y.}\ \bibnamefont
  {Wen}}, \bibinfo {author} {\bibfnamefont {K.-T.}\ \bibnamefont {Lin}},
  \bibinfo {author} {\bibfnamefont {A.~F.}\ \bibnamefont {Kockum}}, \bibinfo
  {author} {\bibfnamefont {B.}~\bibnamefont {Suri}}, \bibinfo {author}
  {\bibfnamefont {H.}~\bibnamefont {Ian}}, \bibinfo {author} {\bibfnamefont
  {J.~C.}\ \bibnamefont {Chen}}, \bibinfo {author} {\bibfnamefont {S.~Y.}\
  \bibnamefont {Mao}}, \bibinfo {author} {\bibfnamefont {C.~C.}\ \bibnamefont
  {Chiu}}, \bibinfo {author} {\bibfnamefont {P.}~\bibnamefont {Delsing}},
  \bibinfo {author} {\bibfnamefont {F.}~\bibnamefont {Nori}}, \bibinfo {author}
  {\bibfnamefont {G.-D.}\ \bibnamefont {Lin}}, \ and\ \bibinfo {author}
  {\bibfnamefont {I.-C.}\ \bibnamefont {Hoi}},\ }\href {\doibase
  10.1103/PhysRevLett.123.233602} {\bibfield  {journal} {\bibinfo  {journal}
  {Phys. Rev. Lett.}\ }\textbf {\bibinfo {volume} {123}},\ \bibinfo {pages}
  {233602} (\bibinfo {year} {2019})}\BibitemShut {NoStop}%
\end{thebibliography}%

\end{document}